\documentclass[aps,pra,reprint]{revtex4-2}
\usepackage{graphicx, amsmath, amssymb, hyperref, multirow, xcolor}
\usepackage[caption=false]{subfig}
\allowdisplaybreaks[1] 

\begin{document}


\newcommand{\comb}[2]{{\begin{pmatrix} #1 \\ #2 \end{pmatrix}}}
\newcommand{\braket}[2]{{\left\langle #1 \middle| #2 \right\rangle}}
\newcommand{\bra}[1]{{\left\langle #1 \right|}}
\newcommand{\ket}[1]{{\left| #1 \right\rangle}}
\newcommand{\ketbra}[2]{{\left| #1 \middle\rangle \middle \langle #2 \right|}}

\newcommand{\fref}[1]{Fig.~\ref{#1}}
\newcommand{\Fref}[1]{Figure~\ref{#1}}
\newcommand{\sref}[1]{Sec.~\ref{#1}}
\newcommand{\Sref}[1]{Section~\ref{#1}}
\newcommand{\subsref}[1]{Subsec.~\ref{#1}}
\newcommand{\subsubsref}[1]{Subsubsec.~\ref{#1}}
\newcommand{\tref}[1]{Table~\ref{#1}}


\title{Spatial Search by Nonlinear Quantum Walk}

\author{David A. Meyer}
	\affiliation{Department of Mathematics, University of California, San Diego, La Jolla, CA 92093-0112}
	\email{dmeyer@ucsd.edu}
\author{Thomas G. Wong}
	\affiliation{Department of Physics, Creighton University, 2500 California Plaza, Omaha, NE 68178}
	\email{thomaswong@creighton.edu}

\begin{abstract}
	Many-body quantum systems with effective nonlinearities have been shown to speed up quantum search on the complete graph, \textit{i.e.}, the combinatorial version of Grover's algorithm, at the expense of the number of particles needed for the effective nonlinearity to hold. Physically, however, data may not be arranged in an all-to-all network, and the task of searching incomplete graphs is the spatial search problem. We explore spatial search using a continuous-time nonlinear quantum walk on a variety of graphs. First, we consider incomplete graphs that are ``sufficiently complete'' so as to asymptotically search like the complete graph under a continuous-time (linear) quantum walk, which includes strongly regular graphs such as Paley graphs, regular graphs such as hypercubes, and irregular graphs such as complete bipartite graphs. For these sufficiently complete graphs, we analytically prove nonlinear speedups for Paley graphs and for complete bipartite graphs whose two partite sets both have size $\Theta(N)$, for suitable cubic and cubic-quintic nonlinearities, and we give numerical evidence for stronger nonlinearities and for hypercubes.  Second, we explore arbitrary-dimensional cubic lattices, and we numerically show that certain nonlinearities speed up search on sufficiently high dimensional lattices. Thus, nonlinear quantum search can remain viable even when the underlying graph is incomplete.
\end{abstract}

\maketitle


\section{Introduction}

\begin{figure*}
\begin{center}
	\subfloat[]{
		\includegraphics{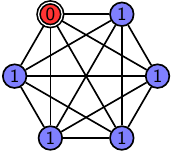}
		\label{fig:complete}
	} \quad
	\subfloat[]{
		\includegraphics{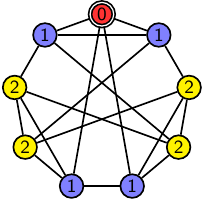}
		\label{fig:paley}
	} \quad
	\subfloat[]{
		\includegraphics{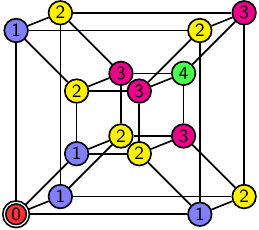}
		\label{fig:hypercube}
	} \quad
	\subfloat[]{
		\includegraphics{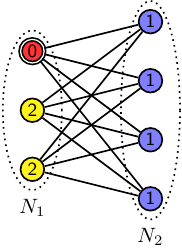}
		\label{fig:bipartite}
	}

	\caption{(a) A complete graph of $N = 6$ vertices. Subfigures (b) through (d) are sufficiently complete graphs, with (b) a Paley graph of $N = 9$ vertices, (c) a 4-dimensional hypercube with $2^4 = 16$ vertices, and (d) a complete bipartite graph with $N_1 = 3$ and $N_2 = 4$ vertices in each partite set. In all, a vertex is marked by an oracle, indicated by a double circle. Identically evolving vertices are identically colored and numerically labeled.}
\end{center}
\end{figure*}

Although Grover's quantum search algorithm \cite{Grover1996} was originally proposed as a digital algorithm, where the state of the system evolves in discrete-time when acted upon by quantum gates, it can also be formulated as an analog algorithm, where the system evolves in continuous-time by Schr\"odinger's equation
\begin{equation}
	\label{eq:Schrodinger}
	i \hbar \frac{d}{dt} \ket{\psi(t)} = H_0 \ket{\psi(t)},
\end{equation}
with some Hamiltonian $H_0$, and throughout this paper, we set $\hbar = 1$. This continuous-time analogue of Grover's algorithm was first proposed by Farhi and Gutmann \cite{FG1998a}, and Childs and Goldstone \cite{CG2004} later provided a physical, intuitive way of interpreting it as a particle undergoing a continuous-time quantum walk on a complete graph with $N$ vertices, an example of which is shown in \fref{fig:complete}.

The $O(\sqrt{N})$ runtime of Grover's algorithm is optimal \cite{Boyer1998} for the unstructured search problem, assuming that the oracle is queried sequentially. A quantum computer with parallel queries (or equivalently, multiple oracles), however, can search faster than this, as the product of the runtime $T$ and the square of the number of oracles $S$ is lower-bounded by $N$, \textit{i.e.}, $ST^2 = \Omega(N)$ \cite{Zalka1999,Wong3}. Many-body quantum systems undergoing a quantum walk naturally fit this regime, as each of the particles interacts with the oracle at the marked vertex, and so it is equivalent to having parallel queries. Then, it is possible for them to search faster than $O(\sqrt{N})$, at the expense of the number of particles in the many-body system \cite{Wong3,Wong4,Wong40}. Many-body quantum systems, especially Bose-Einstein condensates (BECs), are often asymptotically described by effectively nonlinear Schr\"odinger equations of the form
\begin{equation}
	\label{eq:NLSE}
	i \hbar \frac{\partial}{\partial t} \psi(\mathbf{r},t) = \left[ H_0 - g f\!\left( |\psi(\mathbf{r},t)|^2 \right) \right] \psi(\mathbf{r},t),
\end{equation}
where $g$ is a real coefficient, and $f$ is a real-valued function. For example, when $f(p) = p$, \eqref{eq:NLSE} is the Gross-Pitaevskii equation \cite{Gross1961,Pitaevskii1961} with a cubic nonlinearity that describes BECs with two-body interactions in the mean-field limit \cite{Bose1924,Einstein1924,Einstein1925}. When $f(p) = p - p^2$, \eqref{eq:NLSE} contains a cubic-quintic nonlinearity and describes BECs with both two- and three-body interactions \cite{Gammal2000} or BECs confined in pseudo-1D potentials \cite{Trallero2013}, and it also describes nonlinear Kerr media with defocusing corrections \cite{Kerr1877,Kerr1878,Weinberger2008}. When $f(p) = \ln p$, \eqref{eq:NLSE} contains a loglinear nonlinearity and describes Bose liquids \cite{Avdeenkov2011}.

While each of these nonlinearities can speed up search on the complete graph \cite{Wong4,Wong40}, in practice, physically implementing a search algorithm in space could require that the graph be far from complete, such as a planar graph \cite{Benioff2002,AA2005}. This raises the question of whether the nonlinear Schr\"odinger equation \eqref{eq:NLSE} speeds up search on such ``spatial search'' problems, where the graph is incomplete. Some related work considered nonlinear search on the 2D grid using the Childs–Ge crystal-lattice Hamiltonian with a four-site unit cell, linear (Dirac-point) dispersion, and a modified oracle that disconnects the marked vertex from its neighbors~\cite{DiMolfetta2020}. In contrast, here we study nonlinear search using standard continuous-time quantum walks driven by the graph Laplacian.

In this paper, we explore spatial search by nonlinear quantum walk on two broad categories of graphs. The first kind, which we explore in \sref{sec:sufficiently-complete}, are graphs that are not complete, yet are ``sufficiently complete'' in that they solve the linear search problem in a similar manner as the complete graph, asymptotically. We will analytically prove such speedups for Paley graphs and for complete bipartite graphs whose two partite sets both have size $\Theta(N)$, for modest cubic and cubic-quintic nonlinearities, and we will give numerical evidence for stronger nonlinearities and for hypercubes.  The second type of graph is $d$-dimensional lattices with equal length sides, which we investigate in \sref{sec:lattices}. We will numerically show that sufficiently high dimensional lattices support spatial search with nonlinearities of certain forms. Altogether, this demonstrates that nonlinear quantum walks can speed up the spatial search problem. We will conclude in \sref{sec:conclusion}.


\section{\label{sec:sufficiently-complete}Sufficiently Complete Graphs}

In this section, we explore search on sufficiently complete graphs. In \subsref{subsec:sufficiently-complete-definition}, we will define sufficiently complete graphs as graphs on which a linear quantum walk asymptotically searches like the complete graph. Three examples will be explored, namely Paley graphs, hypercubes, and complete bipartite graphs. Then, in \subsref{subsec:sufficiently-complete-nonlinear}, we will explore nonlinear quantum search on them with cubic, cubic-quintic, and loglinear nonlinearities in \subsubsref{subsubsec:sufficient-cubic}, \ref{subsubsec:sufficient-cubic-quintic}, and \ref{subsubsec:sufficient-loglinear}, respectively.  We will analytically prove that Paley graphs and complete bipartite graphs with $N_1 = \Theta(N)$ and $N_2 = \Theta(N)$ support nonlinear quantum search with modest cubic and cubic-quintic nonlinearities.  We will also provide numerical evidence for hypercubes and for stronger nonlinearities, as well as numerical evidence that Paley and complete bipartite graphs support the loglinear nonlinearity.


\subsection{\label{subsec:sufficiently-complete-definition}Definition via Linear Search}

We define \emph{sufficiently complete} graphs as graphs on which a continuous-time (linear) quantum walk asymptotically (\textit{i.e.}, for large numbers of vertices) searches like the complete graph. To make this mathematically precise, let us review search on the complete graph \cite{CG2004,Wong35}.

\begin{figure*}
\begin{center}
	\subfloat[]{
		\includegraphics[width=1.65in]{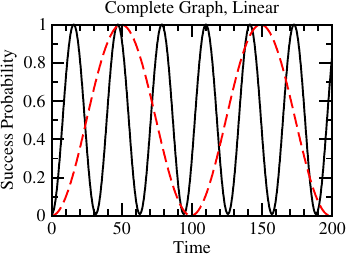}
		\label{fig:complete-linear}
	}
	\subfloat[]{
		\includegraphics[width=1.65in]{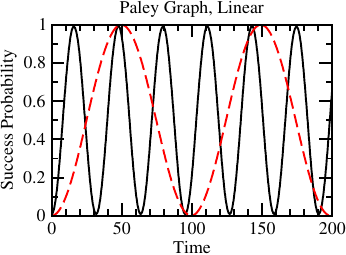}
		\label{fig:paley-linear}
	}
	\subfloat[]{
		\includegraphics[width=1.65in]{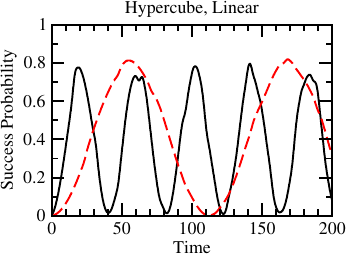}
		\label{fig:hypercube-linear}
	}
	\subfloat[]{
		\includegraphics[width=1.65in]{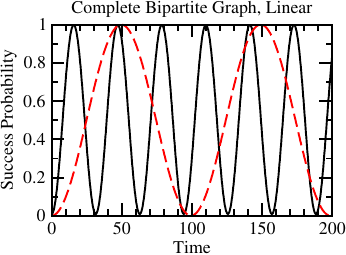}
		\label{fig:bipartite-linear}
	}

	\subfloat[]{
		\includegraphics[width=1.65in]{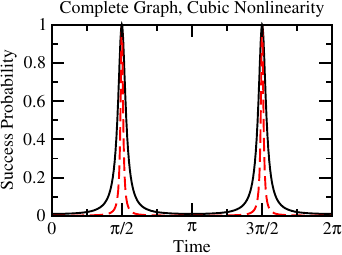}
		\label{fig:complete-cubic}
	}
	\subfloat[]{
		\includegraphics[width=1.65in]{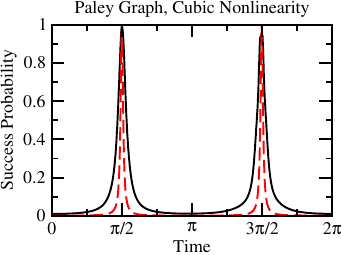}
		\label{fig:paley-cubic}
	}
	\subfloat[]{
		\includegraphics[width=1.65in]{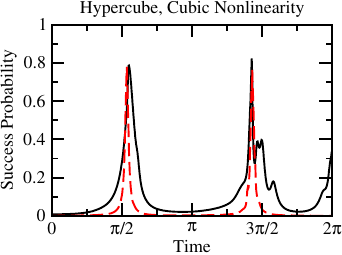}
		\label{fig:hypercube-cubic}
	}
	\subfloat[]{
		\includegraphics[width=1.65in]{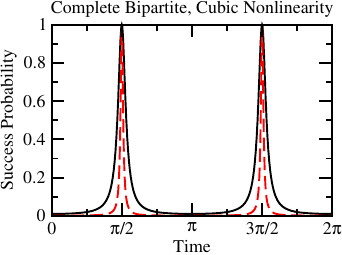}
		\label{fig:bipartite-cubic}
	}

	\subfloat[]{
		\includegraphics[width=1.65in]{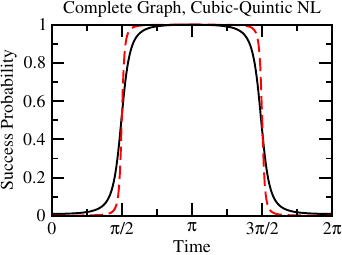}
		\label{fig:complete-cubic-quintic}
	}
	\subfloat[]{
		\includegraphics[width=1.65in]{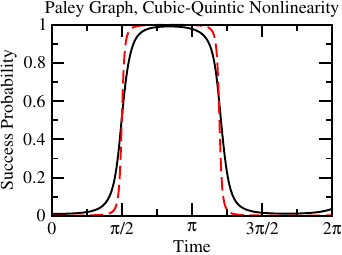}
		\label{fig:paley-cubic-quintic}
	}
	\subfloat[]{
		\includegraphics[width=1.65in]{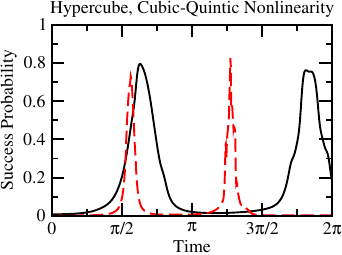}
		\label{fig:hypercube-cubic-quintic}
	}
	\subfloat[]{
		\includegraphics[width=1.65in]{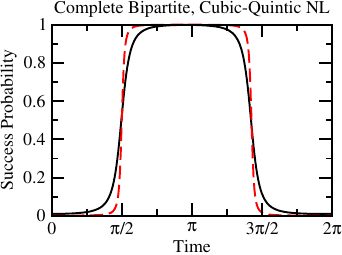}
		\label{fig:bipartite-cubic-quintic}
	}

	\subfloat[]{
		\includegraphics[width=1.65in]{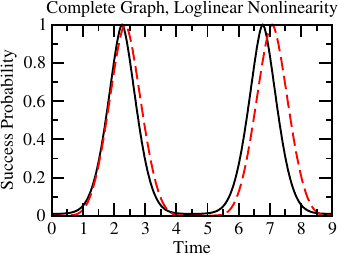}
		\label{fig:complete-loglinear}
	}
	\subfloat[]{
		\includegraphics[width=1.65in]{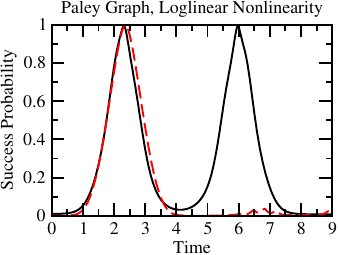}
		\label{fig:paley-loglinear}
	}
	\subfloat[]{
		\includegraphics[width=1.65in]{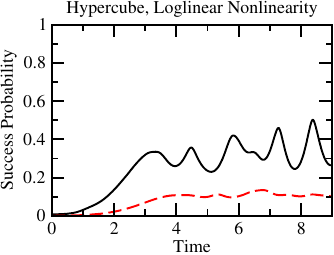}
		\label{fig:hypercube-loglinear}
	}
	\subfloat[]{
		\includegraphics[width=1.65in]{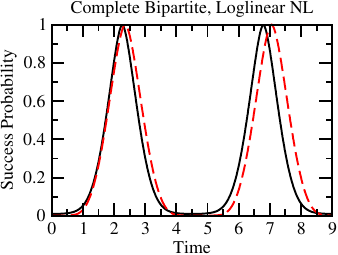}
		\label{fig:bipartite-loglinear}
	}
	\caption{\label{fig:sufficient}Success probability as a function of time for quantum search on various graphs with various nonlinearities (NLs) using $\gamma_c(t)$ \eqref{eq:gammac} with $\ell = 1$. The first row [subfigures (a)-(d)] is the linear ($g = 0$) algorithm, the second row [subfigures (e)-(h)] is the cubic nonlinearity $f(p) = p$ with $g = N-1$, the third row [subfigures (i)-(l)] is the cubic-quintic nonlinearity $f(p) = p-p^2$ with $g = N-1$, and the last row [subfigures (m)-(p)] is the loglinear nonlinearity $f(p) = \ln p$ with $g = \sqrt{N}/\ln N$. The first column [subfigures (a), (e), (i), and (m)] is search on the complete graph with $N = 100$ and $N = 1000$ vertices, the second column [subfigures (b), (f), (j), and (n)] is search on the Paley graph with $N = 101$ and $N = 1001$ vertices, the third column [subfigures (c), (g), (k), and (o)] is search on the hypercube with $N = 2^7 = 128$ and $N = 2^{10} = 1024$ vertices, and the last column [subfigures (d), (h), (l), and (p)] is search on the complete bipartite graph with $(N_1,N_2) = (80,20)$ and $(800,200)$. In each, the solid black curve corresponds to the graph with fewer vertices, and the dashed red curve corresponds to the graph with more vertices.}
\end{center}
\end{figure*}

Labeling the $N$ vertices of the graph $\{ \ket{1}, \ket{2}, \dots, \ket{N} \}$, the system $\ket{\psi(t)}$ begins in an equal superposition $\ket{s}$ of all the vertices:
\begin{equation}
	\label{eq:psi0}
	\ket{\psi(0)} = \ket{s} = \frac{1}{\sqrt{N}} \sum_{j = 1}^{N} \ket{j}.
\end{equation}
Then, it evolves by Schr\"odinger's equation \eqref{eq:Schrodinger} with Hamiltonian
\begin{equation}
	\label{eq:H0}
	H_0 = -\gamma L - \ketbra{w}{w},
\end{equation}
where $\gamma$ is the amplitude per unit time of the particle jumping to an adjacent vertex, $L = A - D$ is the discrete Laplacian with $A$ the adjacency matrix (an $N$-by-$N$ matrix with $A_{ij} = 1$ if vertices $i$ and $j$ are adjacent and $0$ otherwise) and $D$ the degree matrix (an $N$-by-$N$ matrix with $D_{ii} = \deg(i)$ on the diagonal and $0$ otherwise), and $\ket{w}$ with $w \in \{1, 2, \dots, N\}$ is the ``marked'' vertex that we are searching for.

The system evolves such that all the unmarked vertices evolve identically \cite{Wong10}, as indicated in \fref{fig:complete}, where vertices that evolve identically have the same color and numerical label, with the marked vertex colored red and labeled 0, and the unmarked vertices colored blue and labeled 1. So, the system evolves in a two-dimensional subspace. When the jumping rate takes its ``critical'' value of $\gamma_L = 1/N$ \cite{CG2004}, two eigenvectors and corresponding eigenvalues of $H_0$ are
\begin{equation}
	\label{eq:complete-eigensystem}
	\begin{gathered}
		\ket{\psi_{0,1}} = \frac{1}{\sqrt{2}} \sqrt{\frac{\sqrt{N}}{\sqrt{N} \pm 1}} \left( \ket{s} \pm \ket{w} \right), \\
		E_{0,1} = \mp \frac{1}{\sqrt{N}} \quad \Rightarrow \quad \Delta E = \frac{2}{\sqrt{N}},
	\end{gathered}
\end{equation}
where $\Delta E = E_1 - E_0$ is the difference in eigenvalues, or energy gap. Then, the system evolves from $\ket{s}$ to $\ket{w}$ in time $\pi/\Delta E = \pi\sqrt{N}/2$ \cite{Wong10,Wong10Cor}. This evolution can be seen in \fref{fig:complete-linear}, where we have plotted the success probability (\textit{i.e.}, the probability of finding the particle at the marked vertex if its position were measured) versus time for a complete graph with $N = 100$ vertices and $N = 1000$ vertices as the solid black and dashed red curves, respectively. The black curve reaches a success probability of 1 at time $\pi\sqrt{100}/2 \approx 15.708$, and the dashed red curve reaches a success probability of 1 at time $\pi\sqrt{1000}/2 \approx 49.673$, in agreement with the analytical result. Thus, a continuous-time quantum walk searches the complete graph in $O(\sqrt{N})$ time, just like Grover's algorithm solves the unstructured search problem in $O(\sqrt{N})$ timesteps.

\begin{table*}
	\caption{\label{table:sufficient-linear} Some sufficiently complete graphs, the dimension of the subspace of their exact evolution, the number of vertices of each type (with the single marked vertex corresponding to $m_0$), the critical jumping rate for the linear quantum search algorithm, and the order in which it converges to the complete graph's evolution. Note ${}_nC_k$ denotes the combination ``$n$ choose $k$,'' and the complete bipartite graph assumes the marked vertex is in the partite set with $N_1$ vertices.}
\begin{ruledtabular}
\begin{tabular}{ccccccc}
	\multirow{2}{*}{Graph} & Subspace & \multicolumn{3}{c}{\multirow{2}{*}{Number of Vertices of Each Type}} & \multirow{2}{*}{$\gamma_L$} & \multirow{2}{*}{$\epsilon$} \\
	& Dimension & & \\
	\hline
	Paley Graph & $3$ & $|m_0| = 1$ & $|m_1| = \frac{N-1}{2}$ & $|m_2| = \frac{N-1}{2}$ & $\frac{2}{N-1}$ & $\displaystyle \frac{1}{\sqrt{N}}$ \\
	$n$-dimensional Hypercube & $n+1$ & \multicolumn{3}{c}{$|m_k| = {}_nC_k, \quad k = 0, 1, \dots, n$} & $\displaystyle \frac{1}{2^{n+1}} \sum_{k=1}^{n} \frac{{}_nC_k}{k}$ & $\displaystyle \frac{1}{n}$ \\
	Complete Bipartite Graph & $3$ & $|m_0| = 1$ & $|m_1| = N_2$ & $|m_2| = N_1 - 1$ & $\displaystyle \frac{1}{N_2}$ & $\displaystyle \frac{1}{\sqrt{N}}$ \\
\end{tabular}
\end{ruledtabular}
\end{table*}

Now, a sufficiently complete graph is a graph where, with appropriate choice of $\gamma$, two of the eigenvectors and corresponding eigenvalues of its linear search Hamiltonian $H_0$ \eqref{eq:H0} asymptotically take the form of \eqref{eq:complete-eigensystem}. That is, for large $N$, the two eigenvectors and eigenvalues of $H_0$ take the form
\begin{equation}
	\label{eq:2D-asymptotic}
	\begin{gathered}
		\ket{\psi_{0,1}} \propto \ket{s} \pm \ket{w} + O(\epsilon) \ket{e}, \\
		\Delta E = \frac{2}{\sqrt{N}} \left[ 1 + O(\epsilon) \right],
	\end{gathered}
\end{equation}
for some $\epsilon$ that goes to 0 for large $N$, and some ``extra'' vector $\ket{e}$. That is, even though search on a sufficiently complete graph may exactly occur in some higher-dimensional subspace, for large $N$, it evolves more and more like the complete graph in a two-dimensional subspace.

It remains an open question as to which graphs support fast quantum search \cite{Wong7}, so not all sufficiently complete graphs are known. Nonetheless, some examples of them include strongly regular graphs \cite{Wong5}, tetrahedral graphs \cite{Wong20}, and Erd\H{o}s-R\'{e}nyi random graphs \cite{Chakraborty2015}. In this paper, we will focus on three examples of sufficiently complete graphs, which we detail next.

The first example is that of Paley graphs \cite{Cameron1991}, an example of which is shown in \fref{fig:paley}. They are a family of strongly regular graphs, and for search on all strongly regular graphs, there are only three distinct types of vertices: the marked vertex $w$, vertices adjacent to $w$, and vertices nonadjacent to $w$ \cite{Wong5}. In \fref{fig:paley}, vertices of each type have unique colors and respective labels 0, 1, and 2. If we call these three sets of identically-evolving vertices $m_0$, $m_1$, and $m_2$, then the number of vertices of each type is $|m_0| = 1$ and $|m_1| = |m_2| = (N-1)/2$, as listed in the first row of \tref{table:sufficient-linear}. As shown in \cite{Wong5}, when $\gamma$ takes a critical value equal to the reciprocal of the degree of the graph, two of the eigenstates of the search Hamiltonian $H_0$ \eqref{eq:H0} take the form of \eqref{eq:2D-asymptotic} with $\epsilon = 1/\sqrt{N}$. This is also summarized in the first row of \tref{table:sufficient-linear}. As a check that Paley graphs are indeed sufficiently complete, in \fref{fig:paley-linear}, the success probability of the search algorithm is plotted against time for search on the Paley graph with $N = 101$ and $N = 1001$ vertices, and we see that it closely mimics the evolution of the complete graph in \fref{fig:complete-linear}.

The second example is the $n$-dimensional hypercube, which has $N = 2^n$ vertices. An example of this in four dimensions is shown in \fref{fig:hypercube}. We can label each of the vertices with an $n$-bit string $\ket{z_1 \dots z_n}$. Without loss of generality, we choose the marked vertex to be the string of all zeros, \textit{i.e.}, $\ket{w} = \ket{0 \dots 0}$. Then, the vertices ``one away'' from the marked vertex have bit strings with a single one (\textit{i.e.}, Hamming weight 1), the vertices ``two away'' have bit strings with two ones (\textit{i.e.}, Hamming weight 2), and so forth. In \fref{fig:hypercube}, each vertex is labeled with its Hamming weight. Let us denote the set of vertices with Hamming weight $k$ as $m_k$, which has size $|m_k| = {}_nC_k$ or ``$n$ choose $k$.'' Vertices with the same Hamming weight evolve identically, and since $k = 0, 1, \dots, n$, the system evolves in an $(n+1)$-dimensional subspace. This is summarized in the second row of \tref{table:sufficient-linear}. When the jumping rate takes the critical value $\gamma_L$ shown in the second row of \tref{table:sufficient-linear}, the hypercube is sufficiently complete with $\epsilon = 1/n$ \cite{FGGS2000, CDFGGL2002, CG2004}. Since $n = \log_2 N$, search on the hypercube converges very slowly to search on the complete graph, as reflected in \fref{fig:hypercube-linear}, where we have plotted the success probability for search on the seven- and ten-dimensional hypercubes. We see that the success probability is slightly better when a larger hypercube is used, it is still noticeably less than 1 given the slow convergence. The runtimes also have room for improvement. For example, when $N = 2^{10} = 1024$, the dashed red curve's first peak occurs around $t = 55$, and its second peak occurs around $t = 169$, whereas a complete graph with $N = 1024$ vertices would have its first two peaks at $t = \pi\sqrt{1024}/2 = 50.265$ and $t = 3\pi \sqrt{1024} / 2 \approx 150.80$.

The third example of a sufficiently complete graph is the complete bipartite graph, an example of which is shown in \fref{fig:bipartite}. We denote the number of vertices in the partite sets as $N_1$ and $N_2$, and without loss of generality, we assume the marked vertex in the set with $N_1$ vertices. That is, if the marked vertex is in the partite set with $N_2$ vertices, we can simply swap the labels $N_1$ and $N_2$ so that the marked vertex is now in the set with $N_1$ vertices. Similarly to Paley graphs, the system exactly evolves in a three-dimensional subspace, and the three types of vertices are the marked vertex, vertices adjacent to the marked vertex, and vertices nonadjacent to the marked vertex. Then, the sizes of each set are $|m_0| = 1$, $|m_1| = N_2$, and $|m_2| = N_1 - 1$, as listed in the third row of \tref{table:sufficient-linear}. Vertices of the same type are indicated by identical colors and labels in \fref{fig:bipartite}. As shown in \cite{Wong19}, when the jumping rate takes a critical value of $\gamma_L = 1/N_2$, complete bipartite graphs are sufficiently complete with $\epsilon = 1/\sqrt{N}$. This is reflected in \fref{fig:bipartite-linear}, where we plot the success probability vs time for search on the complete bipartite graph with $(N_1,N_2) = (80,20)$ and $(800,200)$ vertices, and it mimics the complete graph's evolution in \fref{fig:complete-linear}. The results for the complete bipartite graph are summarized in the last row of \tref{table:sufficient-linear}.

Note that sufficiently complete graphs can be regular (meaning each vertex has the same number of neighbors), such as Paley graphs and hypercubes, or they can be irregular, meaning vertices can have different numbers of neighbors, such as the complete bipartite graph. The number of edges in a sufficiently complete graph can also be of the same scale as the complete graph, or it can be significantly less. For example, the complete graph of $N$ vertices has $N(N-1)/2$ edges, and a Paley graph with $N$ vertices has $N(N-1)/4$ edges, which are both $O(N^2)$ edges. On the other hand, a hypercube with $N$ vertices has $N \log_2 (N) / 2$ edges, which is nearly quadratically fewer edges than the complete graph. Similarly, a complete bipartite graph with $N_1$ and $N_2 = N - N_1$ vertices in each partite set has $N_1 N_2$ edges, which takes a minimum of $N-1 = O(N)$ edges when one partite set has a single vertex. This is the minimum number of edges for a connected graph and scales quadratically less than the complete graph's $O(N^2)$, and so a graph can have much fewer edges than the complete graph, but still asymptotically support linear quantum search like the complete graph.

In general, say search on a sufficiently complete graph exactly evolves in an $M$-dimensional subspace, so the $N$ vertices of the graph can be grouped together in $M$ sets $m_k$ of size $|m_k|$, with $k = 0, 1, \dots, M-1$, where all vertices in the same set evolve identically. Then, the equal superpositions of identically evolving vertices,
\[ \ket{m_k} = \frac{1}{\sqrt{|m_k|}} \sum_{j \in m_k} \ket{j}, \]
form an orthonormal basis $\{ \ket{m_0}, \ket{m_1}, \dots, \ket{m_{M-1}} \}$ for the $M$-dimensional subspace. Then, the state $\ket{\psi(t)}$ can be written as a linear combination of these basis states:
\[ \ket{\psi(t)} = \sum_{k=0}^{M-1} c_k(t) \ket{m_k}. \]
The uniform initial state \eqref{eq:psi0} corresponds to $c_k(0) = \sqrt{|m_k|/N}$.

Without loss of generality, we pick the marked vertex to be $\ket{m_0} = \ket{w}$, so $|m_0| = 1$, and as proved in Appendix~\ref{sec:linear-probabilities}, if two eigenvectors of the linear search Hamiltonian $H_0$ \eqref{eq:H0} take the form \eqref{eq:2D-asymptotic}, then the success probability at time $t$ is
\begin{equation}
	\label{eq:c0}
	|c_0(t)|^2 = \frac{1}{N} \cos^2 \left( \frac{t}{\sqrt{N}} \right) + \sin^2 \left( \frac{t}{\sqrt{N}} \right) + O(\epsilon),
\end{equation}
and the probability in the other types of vertices are
\begin{equation}
	\label{eq:cne0}
	\left| c_{k \ne 0}(t) \right|^2 = \frac{|m_k|}{N} \cos^2 \left( \frac{t}{\sqrt{N}} \right) + O(\epsilon).
\end{equation}
Thus, for large $N$, the success probability reaches 1 in time $\pi \sqrt{N} / 2$, as it does for arbitrary $N$ for the complete graph.


\subsection{\label{subsec:sufficiently-complete-nonlinear}Nonlinear Search}

In this subsection, we explore search on sufficiently complete graphs 
with a nonlinear quantum walk governed by nonlinear Schr\"odinger 
equations of the form \eqref{eq:NLSE}, \textit{i.e.},
\[ 
    i \frac{\partial}{\partial t} \psi(\mathbf{r},t) 
    = H(t) \psi(\mathbf{r},t), 
\]
where the effective Hamiltonian is
\[ 
    H(t) = H_0 - V(t), 
\]
and
\[ 
    V(t) = g f(|\psi(\mathbf{r},t)|^2) 
\]
is a nonlinear ``self-potential,'' which for positive $g$ speeds up 
the buildup of amplitude.

In the vertex basis $\{ \ket{1}, \ket{2}, \dots, \ket{N} \}$, the 
self-potential is
\begin{equation}
	\label{eq:v}
	V(t) = g \sum_{j=1}^N f\!\left( \left| \braket{j}{\psi(t)} 
	                               \right|^2 
	                        \right) \ketbra{j}{j}.
\end{equation}
Even with this nonlinearity, the system evolves in the same 
$M$-dimensional subspace as the linear algorithm, which was spanned by 
orthonormal basis vectors 
$\{ \ket{m_0}, \ket{m_1}, \dots, \ket{m_{M-1}} \}$.  In this subspace 
the self-potential is
\[ 
    V(t) = g \sum_{k = 0}^{M-1} f_k(t) \ketbra{m_k}{m_k}, 
\]
where
\begin{equation}
	\label{eq:fk}
	f_k(t) = f\!\left( \frac{|c_k(t)|^2}{|m_k|} \right).
\end{equation}
Here $|c_k(t)|^2$ is the total probability of finding the walker in 
the set of vertices of type $k$, and since there are $|m_k|$ vertices 
of this type, $|c_k(t)|^2/|m_k|$ is the probability of finding the 
walker at one particular vertex of the $k^{\mathrm{th}}$ type.  Then
$f_k(t)$ is this per-vertex probability evaluated by $f$.

Using this expression for $V(t)$ and the linear search Hamiltonian for 
$H_0$~\eqref{eq:H0}, the effective Hamiltonian in the $M$-dimensional 
subspace is
\begin{align*}
	H(t)
		&= H_0 - V(t) \\
		&= -\gamma L - \ketbra{m_0}{m_0} - g \sum_{k = 0}^{M-1} f_k(t) \ketbra{m_k}{m_k}.
\end{align*}
To work toward a critical jumping rate, we pull out of the sum the 
$k = 0$ term and another term, say the $\ell^{\mathrm{th}}$ term with 
$\ell \in \{ 1, 2, \dots, M-1 \}$:
\begin{align*}
	H(t)
		&= -\gamma L - \left[ 1 + g f_0(t) \right] \ketbra{m_0}{m_0} - g f_\ell(t) \ketbra{m_\ell}{m_\ell} \\
		&\quad - g \sum_{k \ne 0,\ell} f_k(t) \ketbra{m_k}{m_k}.
\end{align*}
Now, we add and subtract 
$gf_\ell(t)\sum_{k=0}^{M-1}\ketbra{m_k}{m_k} = gf_\ell(t)\mathbb{I}$, 
where $\mathbb{I}$ is the $M \times M$ identity matrix, resulting in
\begin{align*}
	H(t)
		&= -\gamma L - \left[ 1 + g f_0(t) - g f_\ell(t) \right] \ketbra{m_0}{m_0} \\
		&\quad - g \sum_{k \ne 0,\ell} \left[ f_k(t) - f_\ell(t) \right] \ketbra{m_k}{m_k} - g f_\ell(t) \mathbb{I}.
\end{align*}
The term proportional to the identity matrix can be dropped, as it 
only contributes a global phase, which is unobservable.  Then,
\begin{align*}
	H(t)
		&= -\gamma L - \left[ 1 + g f_0(t) - g f_\ell(t) \right] \ketbra{m_0}{m_0} \\
		&\quad - g \sum_{k \ne 0,\ell} \left[ f_k(t) - f_\ell(t) \right] \ketbra{m_k}{m_k}.
\end{align*}

In the nonlinear search algorithm on the complete graph 
\cite{Wong3,Wong4}, the critical jumping rate was a time-varying 
function that caused the nonlinear evolution to follow the same path 
as the linear evolution, but with rescaled time.  For sufficiently 
complete graphs the same mechanism is present if the last term in 
$H(t)$ has negligible accumulated effect over the search interval.  To 
make this precise, define
\begin{equation}
    \label{eq:h}
	h(t) = 1 + g\left[ f_0(t) - f_\ell(t) \right]
\end{equation}
and
\begin{equation}
    \label{eq:R}
	R(t) = -g \sum_{k \ne 0,\ell} \left[ f_k(t) - f_\ell(t) \right] \ketbra{m_k}{m_k}.
\end{equation}
Then, up to an irrelevant scalar multiple of the identity,
\[
	H(t) = -\gamma L - h(t) \ketbra{m_0}{m_0} + R(t).
\]
This motivates the time-dependent critical jumping rate
\begin{equation}
	\label{eq:gammac}
	\gamma_c(t) = \gamma_L \left[ 1 + g f_0(t) - g f_\ell(t) \right] = \gamma_L h(t),
\end{equation}
where $\gamma_L$ is the linear algorithm's critical $\gamma$.  When 
$\gamma = \gamma_c(t)$, the effective Hamiltonian is
\begin{align*}
	H(t)
		&= h(t) \left( -\gamma_L L - \ketbra{m_0}{m_0} \right) + R(t) \\
		&= h(t) \left. H_0 \right|_{\gamma = \gamma_L} + R(t).
\end{align*}
Thus, the first term is the linear Hamiltonian at its critical 
jumping rate $\gamma_L$ with a time-dependent multiplicative factor, 
while $R(t)$ is the remainder from the other vertex classes.

If $R(t) = 0$, the evolution is exactly the linear critical search 
evolution with rescaled time
\[
	s(t) = \int_0^t h(u)\,\mathrm{d}u.
\]
Let
\[
	s_* = \frac{\pi}{\Delta E} 
	    = \frac{\pi\sqrt{N}}{2}\left[1 + O(\epsilon)\right]
\]
be the first success time of the linear critical search, using the gap 
estimate \eqref{eq:2D-asymptotic}.  Suppose that $h(t)>0$ on the 
search interval, so that $s(t)$ is monotone increasing, and let $T_N$ 
be defined by $s(T_N) = s_*$, i.e.,
\begin{equation}
    \label{eq:scaled-runtime}
    \int_0^{T_N} h(t)\,\mathrm{d}t = s_*.
\end{equation}
Let
\[
	\ket{\phi(t)} = e^{-i \left. H_0 \right|_{\gamma = \gamma_L}s(t)}\ket{s}
\]
be the corresponding rescaled linear state.  We denote the deviation between the nonlinear and linear search algorithms by
\[ 
    \eta(t) = \|\ket{\psi(t)}-\ket{\phi(t)}\|.
\]
Then, the nonlinear algorithm asymptotically follows the rescaled linear algorithm if, for $0 \le t \le T_N$, the following two conditions hold:
\begin{subequations}
\label{eq:conditions}
\begin{gather}
    h(t) > 0, \label{eq:condition-h} \\
    \eta(t) = o(1). \label{eq:condition-eta}
\end{gather}
\end{subequations}
If these hold, then the nonlinear state at time $T_N$ 
differs by $o(1)$ from the linear critical-search state at its first 
success time.  Hence, the nonlinear success probability at $T_N$ is 
$1-o(1)$, and the physical runtime is obtained by solving \eqref{eq:scaled-runtime} for $T_N$:
\begin{equation}
    \label{eq:runtime}
	T_N = \int_0^{s_*} \frac{\mathrm{d}s}{h(t(s))}.
\end{equation}
So, proving a nonlinear speedup requires an estimate showing that this 
integral is smaller than the linear runtime.

To prove the second condition \eqref{eq:condition-eta}, the 
linear trajectory, such as \eqref{eq:c0} and \eqref{eq:cne0}, are 
useful guides, but an analytical proof must ultimately bound \eqref{eq:condition-eta} along the nonlinear trajectory 
itself, often by a perturbative or bootstrap argument.  Such bootstrap 
arguments for nonlinear partial differential equations typically combine Duhamel's formula with Gr\"onwall's 
inequality \cite{Pazy1983,Teschl2012}. Duhamel's formula, also called the variation-of-constants formula, states 
that if $U_A(t,u)$ is the propagator generated by a time-dependent 
operator $A(t)$, then a solution of $i\dot{x}(t) = [A(t)+B(t)]x(t)$ 
with initial condition $x(0)$ satisfies
\[
	x(t) = U_A(t,0)x(0) - i\int_0^t U_A(t,u)B(u)x(u)\,\mathrm{d}u
\]
under the usual hypotheses for finite-dimensional linear 
systems~\cite{Pazy1983,Teschl2012}.  Applying this formula with 
$A(t) = h(t)\left. H_0 \right|_{\gamma=\gamma_L}$ and $B(t) = R(t)$ 
gives
\begin{align*}
	&\ket{\psi(t)} - \ket{\phi(t)} \\
	&\quad =
	-i \int_0^{t}
	e^{-i \left. H_0 \right|_{\gamma = \gamma_L}[s(t)-s(u)]}
	R(u)\ket{\psi(u)}\,\mathrm{d}u .
\end{align*}
Taking norms and using unitarity gives an upper bound on how far apart the states of the nonlinear and linear algorithms are:
\begin{equation}
    \label{eq:Duhamel-bound-t}
	\eta(t)
	\le
	\int_0^{t} \left\|R(u)\right\|\,\mathrm{d}u.
\end{equation}
Since the projectors $\ketbra{m_k}{m_k}$ in $R(t)$ \eqref{eq:R} are mutually orthogonal,
\begin{equation}
    \label{eq:remainder}
	\left\|R(t)\right\|
	=
	g\max_{k\ne0,\ell}\left|f_k(t)-f_\ell(t)\right|.
\end{equation}
Combining \eqref{eq:Duhamel-bound-t} and \eqref{eq:remainder}, a sufficient condition for \eqref{eq:condition-eta} to be satisfied is, for $0 \le t \le T_N$,
\begin{equation}
    \label{eq:condition-remainder}
    \begin{aligned}
    &\sup_{0\le t\le T_N}
    \int_0^{t} \left\| R(u) \right\| \,\mathrm{d}u\\
    &\quad=
    \sup_{0\le t\le T_N}
    \int_0^{t}
    g\max_{k\ne0,\ell}\left|f_k(u)-f_\ell(u)\right|
    \,\mathrm{d}u
    =
    o(1),
    \end{aligned}
\end{equation}
which says that the accumulated effect of the remainder must go to zero uniformly up to the search time.

Equivalently, in the rescaled time variable $s$, the 
accumulated-remainder condition in \eqref{eq:condition-remainder} can be written 
as
\[
	\int_0^{s_*}
	\frac{
	g\max_{k\ne0,\ell}\left|f_k(t(s))-f_\ell(t(s))\right|
	}{
	h(t(s))
	}
	\,\mathrm{d}s
	=
	o(1).
\]
A useful pointwise sufficient condition is therefore
\[
	\frac{
	g\max_{k\ne0,\ell}\left|f_k(t)-f_\ell(t)\right|
	}{h(t)}
	=
	o\!\left(\frac{1}{\sqrt{N}}\right)
\]
uniformly for $0\le s(t)\le s_*$.  This pointwise condition is 
stronger than necessary, but it makes clear that one must control the 
accumulated operator-norm effect of the omitted diagonal term, not 
merely compare signed scalar coefficients instantaneously.

Analyzing when the accumulated-remainder condition \eqref{eq:condition-remainder} holds typically involves Gr\"onwall's 
inequality \cite{Teschl2012} and will depend on the form 
of the nonlinearity and the graph.  In the following subsubsections we 
explore this with the cubic $f(p) = p$, cubic-quintic $f(p) = p-p^2$, and 
loglinear $f(p) = \ln p$ nonlinearities with the Paley graph, 
hypercube, and complete bipartite graph.


\subsubsection{\label{subsubsec:sufficient-cubic}Cubic Nonlinearity}

\begin{table*}
	\caption{\label{table:sufficient-summary}The existence of various 
nonlinear quantum search algorithms on a variety of sufficiently 
complete graphs based on analytical and numerical results.  The 
analytical complete bipartite entries assume $N_1 = \Theta(N)$ and 
$N_2 = \Theta(N)$.  The cubic Paley entry with $g = N-1$ is proved using 
a Paley-specific remainder estimate.  The other entries marked numerical 
are supported by the simulations in \fref{fig:sufficient}; for the 
loglinear nonlinearity on the hypercube, the simulations do not support 
the existence of the search algorithm for $g = \sqrt{N}/\ln N$.}
\begin{ruledtabular}
\begin{tabular}{cccc}
	Nonlinearity & Paley & Hypercube & Complete Bipartite \\
	\colrule
    Cubic            & $g \ll \sqrt{N}$ (analytical)        & \multirow{2}{*}{$g = N-1$ (numerical)} & $g \ll \sqrt{N}$ (analytical) \\
    $f(p) = p$       & $g = N-1$ (analytical and numerical) &                                        & $g = N-1$ (numerical) \\
    \colrule
	Cubic-Quintic    & $g \ll \sqrt{N}$ (analytical)        & \multirow{2}{*}{$g = N-1$ (numerical)} & $g \ll \sqrt{N}$ (analytical) \\
    $f(p) = p - p^2$ & $g = N-1$ (numerical)                &                                        & $g = N-1$ (numerical) \\
    \colrule
	Loglinear        & \multirow{2}{*}{$\displaystyle g = \frac{\sqrt{N}}{\ln N}$ (numerical)} & \multirow{2}{*}{not supported for $\displaystyle g = \frac{\sqrt{N}}{\ln N}$ (numerical)} & \multirow{2}{*}{$\displaystyle g = \frac{\sqrt{N}}{\ln N}$ (numerical)} \\
    $f(p) = \ln(p)$ \\
\end{tabular}
\end{ruledtabular}
\end{table*}

The cubic nonlinearity takes the form $f(p) = p$.  In 
\fref{fig:complete-cubic}, we plot the success probability when 
searching the same complete graphs from \fref{fig:complete-linear}, 
but now with the cubic nonlinearity with $g = N-1$ and $\ell = 1$, and 
we see that the success probability reaches 1 at a constant time of 
$\pi/2$~\cite{Wong3,Wong4}.  Now, let us explore whether sufficiently 
complete graphs mimic this behavior.

To begin, we denote the probability at a vertex of type $k$ by $p_k(t)$. That is,
\[
    p_k(t) = \frac{|c_k(t)|^2}{|m_k|}.
\]
For the cubic nonlinearity, $f_k(t) = f(|c_k(t)|^2/|m_k|) = |c_k(t)|^2/|m_k| = p_k(t)$, and so the norm of the remainder \eqref{eq:remainder} has 
the especially simple form
\begin{equation}
    \label{eq:cubic-remainder}
	\|R(t)\|
	=
    g\max_{k\ne0,\ell}
	\left|
	p_k(t)
	-
	p_\ell(t)
	\right|.
\end{equation}
Similarly, let
\[
  p_k^L(s) = \frac{|c_k^L(s)|^2}{|m_k|}
\]
denote the corresponding per-vertex probability for the linear critical
evolution at rescaled time $s$. 
The linear estimates from \eqref{eq:c0} and \eqref{eq:cne0} give, 
along the linear critical trajectory,
\begin{equation}
	\label{eq:f0}
	p_0^L(s)
	=
	\frac{1}{N}\cos^2\!\left(\frac{s}{\sqrt{N}}\right)
	+
	\sin^2\!\left(\frac{s}{\sqrt{N}}\right)
	+
	O(\epsilon),
\end{equation}
\begin{equation}
    \label{eq:fne0}
    p_k^L(s)
    =
    \frac{1}{N}\cos^2\!\left(\frac{s}{\sqrt{N}}\right)
    +
    O\!\left(\frac{\epsilon}{|m_k|}\right),
    \qquad k\ne0,\ell .
\end{equation}
and
\begin{equation}
	\label{eq:fl}
	p_\ell^L(s)
	=
	\frac{1}{N}\cos^2\!\left(\frac{s}{\sqrt{N}}\right)
	+
	O\!\left(\frac{\epsilon}{|m_\ell|}\right).
\end{equation}
Equations \eqref{eq:fne0} and \eqref{eq:fl} suggest that the unmarked classes are balanced, but 
they are estimates for the linear trajectory.  As discussed earlier, we can turn these into an 
analytical proof for the nonlinear trajectory using a bootstrap argument by using the 
accumulated-remainder condition \eqref{eq:condition-remainder} that arose from Duhamel's formula, along with Gr\"onwall's inequality. We do this in Appendix~\ref{sec:cubic-proof}, and it yields the following result:

Let
\begin{equation}
	\label{eq:cubic-Bell}
	B_\ell(N):=
	\max_{k\ne0,\ell}
	\left(\frac{1}{|m_k|}+\frac{1}{|m_\ell|}\right).
\end{equation}
Suppose that \eqref{eq:f0}--\eqref{eq:fl} hold uniformly for 
$0\le s\le s_*$, and that
\begin{equation}
    \label{eq:conditions-cubic}
    g\epsilon = o(1),
    \quad
    gB_\ell\sqrt{N} = o(1),
    \quad
    g^2\epsilon B_\ell\sqrt{N} = o(1).
\end{equation}
Then Appendix~\ref{sec:cubic-proof} proves that \eqref{eq:conditions}
are satisfied up to the first success time.  Thus
\eqref{eq:conditions-cubic} are sufficient conditions for the cubic
nonlinear evolution to follow the rescaled linear evolution up to the
first success time, with runtime given by \eqref{eq:cubic-runtime-modest}
and success probability $1-o(1)$.  That is, \eqref{eq:conditions-cubic} are sufficient conditions for the cubic nonlinear 
evolution to follow the rescaled linear evolution up to the first 
success time. Also shown in Appendix~\ref{sec:cubic-proof}, the nonlinear algorithm reaches a success probability of $1 - o(1)$ in time
\begin{equation}
	\label{eq:cubic-runtime-modest}
	T_N
	=
	\frac{\pi\sqrt{N}}{2\sqrt{1+g}}\,[1+o(1)].
\end{equation}

For example, for Paley graphs, \tref{table:sufficient-linear} gives 
$\epsilon = 1/\sqrt{N}$, and choosing $\ell = 1$ gives 
$B_\ell = O(1/N)$.  Hence, \eqref{eq:conditions-cubic} reduces to 
$g\ll\sqrt{N}$, and so we have proven that the cubic nonlinear search algorithm works for Paley graphs when $g \ll \sqrt{N}$.  The same conclusion holds for complete bipartite 
graphs with $N_1 = \Theta(N)$ and $N_2 = \Theta(N)$, for which 
$\epsilon = 1/\sqrt{N}$ and $B_\ell = O(1/N)$.  Thus, Paley graphs and 
such complete bipartite graphs analytically support cubic nonlinear 
search for $g\ll\sqrt{N}$, with runtime given by 
\eqref{eq:cubic-runtime-modest}.  The estimates 
\eqref{eq:f0}--\eqref{eq:fl} are too coarse to prove an analogous 
hypercube theorem, because some Hamming-weight classes have small size; 
for the hypercube, our results will be numerical.

We can improve our result for Paley graphs using the same remainder method \eqref{eq:condition-remainder}, and as shown in Appendix~\ref{sec:cubic-paley-proof}, \eqref{eq:conditions} is satisfied when $g = N - 1$, and the success probability reaches $1 - O(N^{-1})$ in constant time. So, cubic nonlinear search on Paley graphs is analytically shown to hold when $g = N - 1$ as well.

Numerically, the second row of \fref{fig:sufficient} shows that all three graphs (Paley graphs, hypercubes, and complete bipartite graphs) support the cubic nonlinear search algorithm when $g = N-1$.

These analytical and numerical results are summarized in the first row of 
\tref{table:sufficient-summary}.


\subsubsection{\label{subsubsec:sufficient-cubic-quintic}Cubic-Quintic Nonlinearity}

For the cubic-quintic nonlinearity, let $f(p) = p-p^2$.
In \fref{fig:complete-cubic-quintic}, we plot the success probability
when searching the same complete graphs as before, but now using the
cubic-quintic nonlinearity with $g = N-1$.  The success probability reaches
$1$ at a constant time of $\pi$~\cite{Wong4}.  The success probability
also forms a broad plateau near $1$, in contrast to the sharper peaks for
the cubic nonlinearity in \fref{fig:complete-cubic}.  We now explore
search on sufficiently complete graphs with the cubic-quintic
nonlinearity.

Recall that
\[
    p_k(t) = \frac{|c_k(t)|^2}{|m_k|}
\]
is the probability at an individual vertex of type $k$, and that
$f_k(t) = f(p_k(t))$.  The analysis should be applied to differences of
$f$, because the remainder term in \eqref{eq:remainder} contains
$f_k(t) - f_\ell(t)$.  For $p,q\in[0,1]$,
\begin{equation}
    \label{eq:cq-difference-identity}
    f(p)-f(q) = (p-q)(1-p-q).
\end{equation}
In particular,
\begin{equation}
    \label{eq:cq-lipschitz}
    |f(p)-f(q)|\le |p-q|,
    \quad 0\le p,q\le 1.
\end{equation}
Thus the cubic-quintic remainder \eqref{eq:remainder} satisfies
\begin{align}
    \|R(t)\|
    &=
    g\max_{k\ne0,\ell}|f(p_k(t))-f(p_\ell(t))| \nonumber \\
    &\le
    g\max_{k\ne0,\ell}|p_k(t)-p_\ell(t)| .
    \label{eq:cq-remainder-bound}
\end{align}
This remainder is bounded by the same quantity that appeared in the cubic case in
\eqref{eq:cubic-remainder}, and so the same results about the existence of the nonlinear algorithm apply. In particular, under the conditions
\eqref{eq:conditions-cubic}, the nonlinear cubic-quintic trajectory stays
$o(1)$-close to the corresponding rescaled linear trajectory up to the
first success time, provided $0\le s(t)\le s_*$ and $t = O(\sqrt N)$. Furthermore, as shown in Appendix~\ref{sec:cubic-quintic-proof}, the nonlinear algorithm reaches a success probability of $1 - o(1)$ at time
\begin{equation}
    \label{eq:cq-runtime-modest}
    T_N
    =
    \frac{\pi\sqrt N}{2\sqrt{1+g/4}}\,[1+o(1)].
\end{equation}

For example, for Paley graphs, \tref{table:sufficient-linear} gives
$\epsilon = 1/\sqrt N$, and choosing $\ell = 1$ gives
$B_\ell = O(1/N)$.  Thus \eqref{eq:conditions-cubic} reduces to
$g\ll\sqrt N$, and $g/N = o(1)$ is automatic.  The same conclusion holds
for complete bipartite graphs with $N_1 = \Theta(N)$ and
$N_2 = \Theta(N)$, for which $\epsilon = 1/\sqrt N$ and
$B_\ell = O(1/N)$.  Hence, Paley graphs and such complete bipartite graphs
analytically support cubic-quintic nonlinear search for
$g\ll\sqrt N$, with runtime given by \eqref{eq:cq-runtime-modest}.
As in the cubic case, the estimates used here are too coarse to prove
an analogous hypercube theorem, because some Hamming-weight classes have
small size.  The hypercube results for the cubic-quintic nonlinearity
are therefore treated as numerical evidence.

Reiterating that our analytical results are sufficient but not necessary
conditions for the existence of the nonlinear algorithm, we next
numerically explore the cubic-quintic nonlinearity with $g = N-1$.  In the
third row of \fref{fig:sufficient}, we plot the success probability for
search using the cubic-quintic nonlinearity under the same conditions as
for the cubic nonlinearity in the second row.  We again see that the
nonlinearity speeds up search on the Paley graphs in
\fref{fig:paley-cubic-quintic} and complete bipartite graphs in
\fref{fig:bipartite-cubic-quintic} in a similar manner as the complete
graph in \fref{fig:complete-cubic-quintic}, including the broad plateau
near success probability $1$.  For the hypercubes considered in
\fref{fig:hypercube-cubic-quintic}, the success probability does not
form the same broad plateau.  These numerical results for $g = N-1$ are
also reflected in the second row of \tref{table:sufficient-summary}.


\subsubsection{\label{subsubsec:sufficient-loglinear}Loglinear Nonlinearity}

For the loglinear nonlinearity, $f(p) = \ln p$.  In \fref{fig:complete-loglinear}, we plot the success probability when searching the same complete graphs as before, but now with the loglinear nonlinearity and with $g = \sqrt{N}/\ln N$, and we see that the success probability reaches 1 at a constant time around $t = 2.3$ \cite{Wong4}. Now, let us explore search on sufficiently complete graphs with the loglinear nonlinearity.

For the loglinear nonlinearity, the remainder term takes a different
form from the cubic and cubic-quintic cases.  Since
\[
    f_k(t) - f_\ell(t)
    =
    \ln p_k(t) - \ln p_\ell(t)
    =
    \ln\!\left(\frac{p_k(t)}{p_\ell(t)}\right),
\]
\eqref{eq:remainder} becomes
\begin{equation}
    \label{eq:loglinear-remainder}
    \|R(t)\|
    =
    g\max_{k\ne0,\ell}
    \left|
    \ln\!\left(\frac{p_k(t)}{p_\ell(t)}\right)
    \right|.
\end{equation}
Similarly, the time-rescaling factor \eqref{eq:h} is
\begin{equation}
    \label{eq:loglinear-h}
    h(t)
    =
    1 + g\ln\!\left(\frac{p_0(t)}{p_\ell(t)}\right).
\end{equation}
Thus, an analytical proof would require showing both that $h(t) > 0$
and that the accumulated logarithmic remainder in
\eqref{eq:condition-remainder} is $o(1)$.

The linear estimates \eqref{eq:fne0} and \eqref{eq:fl} give additive
control of the unmarked probabilities, but the logarithm requires
relative control.  In particular, the estimates
\[
    p_{k\ne0}^L(s)
    =
    \frac1N\cos^2\!\left(\frac{s}{\sqrt N}\right)
    +
    O\!\left(\frac{\epsilon}{|m_k|}\right)
\]
do not by themselves imply useful bounds on
\[
    \ln\!\left(\frac{p_k^L(s)}{p_\ell^L(s)}\right),
\]
especially near the linear success time, where the leading term
$(1/N)\cos^2(s/\sqrt N)$ becomes small.  Thus, unlike the cubic and
cubic-quintic nonlinearities, the estimates used above do not yield a
remainder/bootstrapping proof for the loglinear case.  A proof would
need sharper lower bounds and relative-error estimates for the ratios
$p_k(t)/p_\ell(t)$ along the nonlinear trajectory.  We therefore treat
the loglinear results below as numerical evidence.

In the last row of \fref{fig:sufficient}, we plot the success probability
for search with the loglinear nonlinearity on the same graphs as in the
preceding rows, using $g = \sqrt{N}/\ln N$.  This value of $g$ is
sufficient for constant-time search on the complete graph, as shown in
\fref{fig:complete-loglinear} and in Ref.~\cite{Wong4}.  The simulations
show that search on the Paley graphs in \fref{fig:paley-loglinear} and
the complete bipartite graphs in \fref{fig:bipartite-loglinear} is sped
up in a manner similar to the complete graph in
\fref{fig:complete-loglinear}.  Thus, although we do not have an
analytical proof for the loglinear case, the numerical evidence suggests
that these graph families support the loglinear nonlinear algorithm for
this choice of $g$.  Search on the hypercube in
\fref{fig:hypercube-loglinear} does not evolve in the same way, however,
so the numerical evidence does not support a loglinear nonlinear
algorithm on the hypercube with $g = \sqrt{N}/\ln N$.  These numerical
results are summarized in the last row of
\tref{table:sufficient-summary}.


\section{\label{sec:lattices}Periodic Lattices}

In this section, we explore quantum search on $d$-dimensional periodic cubic lattices with an equal number of vertices on each side. That is, in 2D, it has $\sqrt{N} \times \sqrt{N}$ vertices, and in 3D, has $N^{1/3} \times N^{1/3} \times N^{1/3}$ vertices, and so forth. First, we will review results for the linear search algorithm, and then we will explore nonlinear quantum search.


\subsection{Linear Search}

\begin{table}
	\caption{\label{table:cubiclattices}Single runtimes and success probabilities for search on $d$-dimensional periodic cubic lattices by quantum walk, and the total runtimes with classical repetitions. Table adapted from \cite{Wong7}.}
\begin{ruledtabular}
\begin{tabular}{cccc}
	$d$     & Success Prob       & Single Runtime       & Total Runtime \\
	\colrule
	$2$     & $O[(\log^2 N)/ N]$ & $O(N / \log N)$      & $O(N^2 / \log^3 N)$ \\
	$3$     & $O(1/N^{1/3})$     & $O(N^{2/3})$         & $O(N)$ \\
	$4$     & $O(1 / \log N)$    & $O(\sqrt{N \log N})$ & $O(\sqrt{N} \log^{3/2} N)$ \\
	$\ge 5$ & $O(1)$             & $O(N^{1/2})$         & $O(N^{1/2})$ \\
\end{tabular}
\end{ruledtabular}
\end{table}

Linear search on $d$-dimensional cubic lattices was explored by Childs and Goldstone in \cite{CG2004}. They defined $d$-component vectors $\vec{k}$ with components
\[ k_j = \frac{2\pi m_j}{N^{1/d}},\]
where
\[ m_j = \begin{cases}
	0, \pm 1, \dots, \pm \frac{1}{2} \left( N^{1/d} - 1 \right), & N^{1/d}\text{ odd}, \\
	0, \pm 1, \dots, \pm \frac{1}{2} \left( N^{1/d} - 2 \right), + \frac{1}{2} N^{1/d}, & N^{1/d}\text{ even}.
\end{cases} \]
Then, they showed that the critical jumping rate for the linear algorithm sums over all nonzero $\vec{k}$'s according to
\[ \gamma_L = \frac{1}{N} \sum_{\vec{k} \ne \vec{0}} \frac{1}{\mathcal{E}(\vec{k})}, \]
where
\[ \mathcal{E}(\vec{k}) = 2 \left[ d - \sum_{j=1}^d \cos(k_j) \right]. \]
With this choice of $\gamma_L$, the linear search algorithm achieves various success probabilities at various runtimes, as summarized in \tref{table:cubiclattices}. For example, in 2D, the success probability reaches $O[(\log^2 N)/ N]$ at time $O(N / \log N)$, and since the success probability is not 1, we expect to repeat the algorithm several times before finding the marked vertex. Dividing the single runtime by the success probability, the expected total runtime with classical repetitions is $O(N^2 / \log^3 N)$. This is much slower than the $O(\sqrt{N})$ runtime for search on sufficiently complete graphs. For comparison, the 2D lattice has $2N = O(N)$ edges, and the complete bipartite graph can have as little as $N-1$ edges, yet search on the complete bipartite graph is much faster. So, the number of edges is not a reliable indicator of whether fast quantum search is possible.

\begin{figure*}
\begin{center}
	\subfloat[]{
		\includegraphics[width=1.65in]{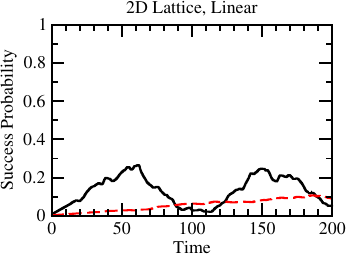}
		\label{fig:lattice-2D-linear}
	}
	\subfloat[]{
		\includegraphics[width=1.65in]{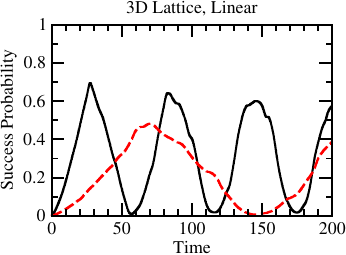}
		\label{fig:lattice-3D-linear}
	}
	\subfloat[]{
		\includegraphics[width=1.65in]{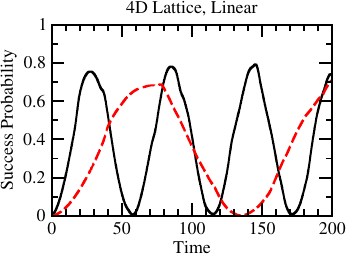}
		\label{fig:lattice-4D-linear}
	}
	\subfloat[]{
		\includegraphics[width=1.65in]{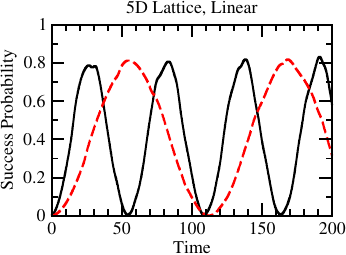}
		\label{fig:lattice-5D-linear}
	}

	\subfloat[]{
		\includegraphics[width=1.65in]{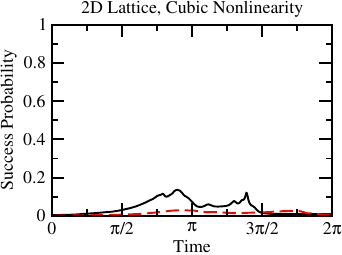}
		\label{fig:lattice-2D-cubic}
	}
	\subfloat[]{
		\includegraphics[width=1.65in]{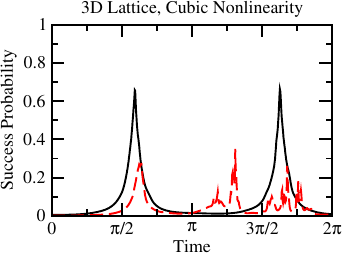}
		\label{fig:lattice-3D-cubic}
	}
	\subfloat[]{
		\includegraphics[width=1.65in]{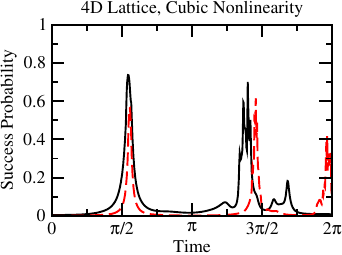}
		\label{fig:lattice-4D-cubic}
	}
	\subfloat[]{
		\includegraphics[width=1.65in]{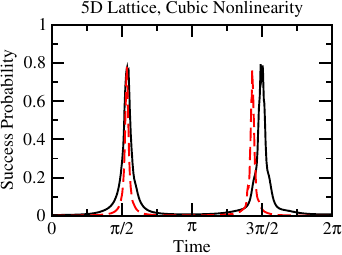}
		\label{fig:lattice-5D-cubic}
	}

	\subfloat[]{
		\includegraphics[width=1.65in]{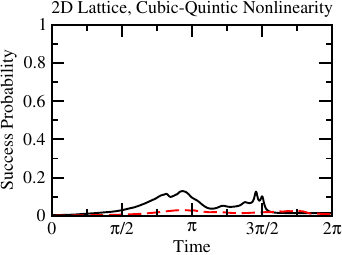}
		\label{fig:lattice-2D-cubic-quintic}
	}
	\subfloat[]{
		\includegraphics[width=1.65in]{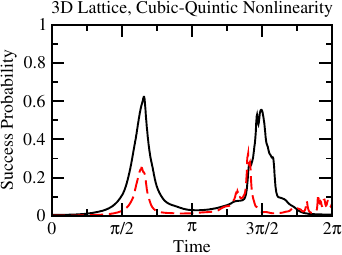}
		\label{fig:lattice-3D-cubic-quintic}
	}
	\subfloat[]{
		\includegraphics[width=1.65in]{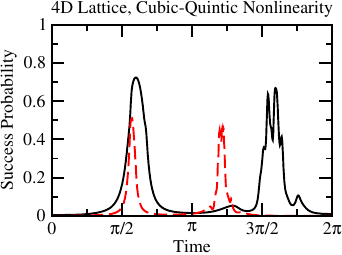}
		\label{fig:lattice-4D-cubic-quintic}
	}
	\subfloat[]{
		\includegraphics[width=1.65in]{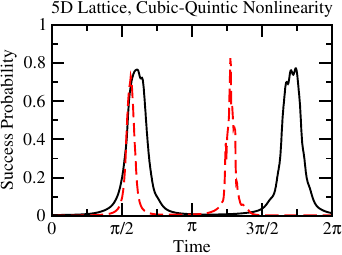}
		\label{fig:lattice-5D-cubic-quintic}
	}

	\subfloat[]{
		\includegraphics[width=1.65in]{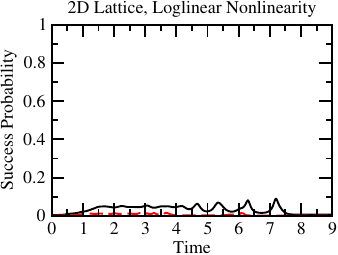}
		\label{fig:lattice-2D-loglinear}
	}
	\subfloat[]{
		\includegraphics[width=1.65in]{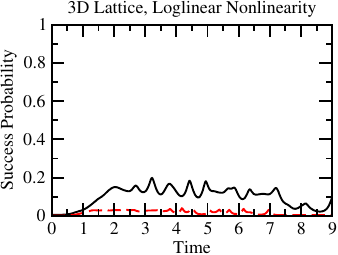}
		\label{fig:lattice-3D-loglinear}
	}
	\subfloat[]{
		\includegraphics[width=1.65in]{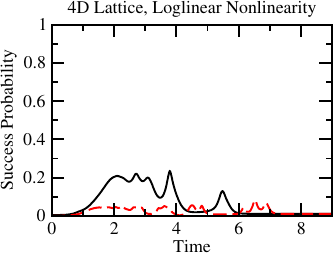}
		\label{fig:lattice-4D-loglinear}
	}
	\subfloat[]{
		\includegraphics[width=1.65in]{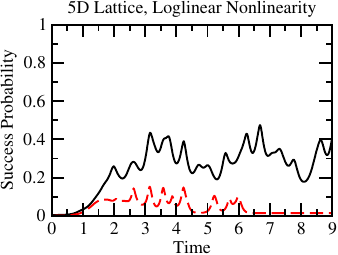}
		\label{fig:lattice-5D-loglinear}
	}
	\caption{\label{fig:lattice}Success probability as a function of time for quantum search on various graphs with various nonlinearities using $\gamma_c(t)$ \eqref{eq:gammac} with $\ell$ chosen to be the vertices adjacent to the marked vertex. The first row [subfigures (a)-(d)] is the linear ($g = 0$) algorithm, the second row [subfigures (e)-(h)] is the cubic nonlinearity $f(p) = p$ with $g = N-1$, the third row [subfigures (i)-(l)] is the cubic-quintic nonlinearity $f(p) = p-p^2$ with $g = N-1$, and the last row [subfigures (m)-(p)] is the loglinear nonlinearity $f(p) = \ln p$ with $g = \sqrt{N}/\ln N$. The first column [subfigures (a), (e), (i), and (m)] is search on the 2D lattice with $N = 16^2 = 256$ and $32^2 = 1024$ vertices, the second column [subfigures (b), (f), (j), and (n)] is search on the 3D lattice with $N = 6^3 = 216$ and $10^3 = 1000$ vertices, the third column [subfigures (c), (g), (k), and (o)] is search on the 4D lattice with $N = 4^4 = 256$ and $6^4 = 1296$ vertices, and the last column [subfigures (d), (h), (l), and (p)] is search on the 5D lattice with $N = 3^5 = 243$ and $4^5 = 1024$ vertices. In each, the solid black curve corresponds to the graph with fewer vertices, and the dashed red curve corresponds to the graph with more vertices.}
\end{center}
\end{figure*}

The success probability for linear search on various 2D, 3D, 4D, and 5D lattices is shown in the first row of \fref{fig:lattice}, and we see that as the dimension increases, the success probability improves, and numerically, the runtime is closer to the complete graph's runtime of $\pi\sqrt{N}/2$.

Note that for fixed dimensions, the lattice is not a sufficiently complete graph. For example, for the 5D lattice, increasing the number of vertices does not cause the success probability to converge to $1$, as it would for a sufficiently complete graph.


\subsection{Nonlinear Search}

\begin{table}
	\caption{\label{table:lattice-summary}Whether various nonlinear quantum search algorithms are supported on various dimension lattices, all supported by numerical simulation.}
\begin{ruledtabular}
\begin{tabular}{cccccc}
	Nonlinearity & $g$ & 2D & 3D & 4D & 5D \\
	\colrule
	Cubic & \multirow{2}{*}{$N - 1$} & \multirow{2}{*}{No} & \multirow{2}{*}{No} & \multirow{2}{*}{No} & \multirow{2}{*}{Yes} \\
    $f(p) = p$ \\
    \colrule
	Cubic-Quintic & \multirow{2}{*}{$N - 1$} & \multirow{2}{*}{No} & \multirow{2}{*}{No} & \multirow{2}{*}{No} & \multirow{2}{*}{Yes} \\
    $f(p) = p - p^2$ \\
    \colrule
	Loglinear & \multirow{2}{*}{$\displaystyle \frac{\sqrt{N}}{\ln N}$} & \multirow{2}{*}{No} & \multirow{2}{*}{No} & \multirow{2}{*}{No} & \multirow{2}{*}{No} \\
    $f(p) = \ln(p)$ \\
\end{tabular}
\end{ruledtabular}
\end{table}

Although $d$-dimensional lattices are not sufficiently complete graphs, we numerically explore search on them using the critical jumping rate from \eqref{eq:gammac} as if they were sufficiently complete, and we choose $\ell$ to correspond to the vertices adjacent to the marked vertex, which all evolve identically. The results are shown in \fref{fig:lattice}, and for each subfigure, the choices of nonlinearity $f(p)$ and nonlinearity coefficient $g$ are identical to those from \fref{fig:sufficient}, \textit{i.e.}, $g = N-1$ for the cubic and cubic-quintic nonlinearities, and $g = \sqrt{N}/\ln N$ for the loglinear nonlinearity.

To gauge the soundness of the nonlinearity algorithms, we compare the height of the success probabilities in the linear algorithm with the nonlinear ones. Starting with the first column, \fref{fig:lattice-2D-linear} shows linear search on 2D lattices. The black curve (corresponding to a 2D lattice with 256 vertices) roughly reaches a success probability of 0.26, while the red curve (corresponding to a 2D lattice with 1024 vertices) roughly reaches a success probability of 0.11. With the cubic nonlinearity in \fref{fig:lattice-2D-cubic}, the success probabilities are substantially lower than the linear case from \fref{fig:lattice-2D-linear}. So, the cubic nonlinear algorithm is failing to evolve as a sped-up version of the linear algorithm, since it does not reach the same success probability. The same holds true for the cubic-quintic nonlinearity in \fref{fig:lattice-2D-cubic-quintic}, and while this graph is curiously similar to the \fref{fig:lattice-2D-cubic}, a close examination of the numbers show that there are slight differences unobservable to the eye. Finishing the first column, \fref{fig:lattice-2D-loglinear} shows that our loglinear nonlinear algorithm is not supported on the 2D lattice, either. This is summarized in the column labeled ``2D'' in \tref{table:lattice-summary}.

The second and third columns of \fref{fig:lattice} depict search on the 3D and 4D lattices. While the cubic and cubic-quintic nonlinearities provide a speedup to a near constant runtime, the decline in success probability compared to the linear case, particularly for the dashed red curve corresponding to the larger lattice, shows that the nonlinear algorithms will asymptotically fail. The poor evolution of the loglinear nonlinearity also shows that 3D and 4D lattices do not support fast nonlinear quantum search. These results are summarized in the ``3D'' and ``4D'' columns of \tref{table:lattice-summary}.

For the last column of \fref{fig:lattice}, \fref{fig:lattice-5D-cubic} and \fref{fig:lattice-5D-cubic-quintic} indicate that 5D lattices do support the cubic and cubic-quintic nonlinearities, as the runtimes are sped up to near constant time, and the success probability remains high like the linear case in \fref{fig:lattice-5D-linear}. As with the other lattices, \fref{fig:lattice-5D-loglinear} shows that the loglinear nonlinearity evolves poorly. These results are summarized in the ``5D'' column of \tref{table:lattice-summary}.

Altogether, our numerical results from \tref{table:lattice-summary} show that nonlinear quantum search is possible on some lattices with some nonlinearities when using the critical jumping rate from \eqref{eq:gammac}.


\section{\label{sec:conclusion}Conclusion}

We have explored whether a nonlinear quantum walk can be used to speed up spatial search. We considered two classes of graphs.  First, we examined graphs on which the linear quantum walk search algorithm asymptotically behaves as it does on the complete graph, which we called sufficiently complete graphs.  We proposed a time-varying critical jumping rate and analytically proved that Paley graphs and complete bipartite graphs with $N_1 = \Theta(N)$ and $N_2 = \Theta(N)$ support search with modest cubic and cubic-quintic nonlinearities.  We also proved a stronger Paley-graph result for the cubic nonlinearity with $g = N-1$ and numerical evidence shows that stronger nonlinearities work on additional sufficiently complete graphs as well.  In addition, we provided numerical evidence that Paley graphs and complete bipartite graphs support search with the loglinear nonlinearity.  Second, we considered search on $d$-dimensional cubic lattices using the same critical jumping rate, and numerical evidence suggests that when $d\ge5$, the lattices support nonlinear quantum search with the cubic and cubic-quintic nonlinearities.

While our results show the existence of nonlinear spatial search algorithms on various graphs, they do not exclude such algorithms in situations where our particular choice of critical jumping rate was unsuccessful.  That is, it could be that a different critical jumping rate can support nonlinear quantum search in situations where ours failed. Furthermore, our approach was restricted to a particular type of continuous-time quantum walk. Other models of quantum computation could be more successful for searching with various graphs or nonlinearities. These are potential topics of further research.


\begin{acknowledgments}
	This material is based upon work supported in part by the National Science Foundation EPSCoR Cooperative Agreement OIA-2044049, Nebraska’s EQUATE collaboration. Any opinions, findings, and conclusions or recommendations expressed in this material are those of the author(s) and do not necessarily reflect the views of the National Science Foundation.
\end{acknowledgments}


\appendix

\section{\label{sec:linear-probabilities}Linear Evolution of Sufficiently Complete Graphs}

In this appendix, for linear search on sufficiently complete graphs, we calculate the probability of finding the walker at the $k$-type vertices, \textit{i.e.}, $|c_k(t)|^2 = |\braket{m_k}{\psi(t)}|^2$.

First, we normalize the eigenvectors $\ket{\psi_{0,1}} = \ket{s} \pm \ket{w} + O(\epsilon) \ket{e}$ from \eqref{eq:2D-asymptotic} by finding their norm:
\begin{widetext}
\begin{align*}
	\braket{\psi_{0,1}}{\psi_{0,1}}
	&= \left[ \bra{s} \pm \bra{w} + O(\epsilon) \bra{e} \right] \left[ \ket{s} \pm \ket{w} + O(\epsilon) \ket{e} \right] \\
	&= \underbrace{\braket{s}{s}}_1 + \underbrace{\braket{w}{w}}_1 + O(\epsilon^2) \underbrace{\braket{e}{e}}_1 \pm ( \underbrace{\braket{s}{w}}_{1/\sqrt{N}} + \underbrace{\braket{w}{s}}_{1/\sqrt{N}} ) + O(\epsilon) \underbrace{( \braket{s}{e} + \braket{e}{s} )}_{\le 2} {}\pm O(\epsilon) \underbrace{( \braket{w}{e} + \braket{e}{w} )}_{\le 2} \\
	&= 2 + O(\epsilon^2) \pm \frac{2}{\sqrt{N}} + O(\epsilon) \pm O(\epsilon) \\
	&= 2 \frac{\sqrt{N} \pm 1}{\sqrt{N}} + O(\epsilon),
\end{align*}
where the inequalities are because for a complex number $z = a+bi$ of length $\le 1$, $z+z^* = (a+bi)+(a-bi)=2a \le 2$. Dividing the eigenvectors by the square root of this norm, the normalized eigenvectors are
\[ \ket{\psi_{0,1}} = \frac{1}{\sqrt{2 \frac{\sqrt{N} \pm 1}{\sqrt{N}} + O(\epsilon)}} \left[ \ket{s} \pm \ket{w} + O(\epsilon) \ket{e} \right]. \]

Now, we can find the initial state $\ket{\psi(0)} = \ket{s}$ in terms of $\ket{\psi_{0,1}}$ and $\ket{e}$. Multiplying both sides of the previous equation by the square root,
\[ \sqrt{2 \frac{\sqrt{N} \pm 1}{\sqrt{N}} + O(\epsilon)} \ket{\psi_{0,1}} = \ket{s} \pm \ket{w} + O(\epsilon) \ket{e} . \]
The above is two equations, one with $\ket{\psi_0}$ and $+$, and the other with $\ket{\psi_1}$ and $-$. Adding the two equations together,
\[ \sqrt{2 \frac{\sqrt{N} + 1}{\sqrt{N}} + O(\epsilon)} \ket{\psi_0} + \sqrt{2 \frac{\sqrt{N} - 1}{\sqrt{N}} + O(\epsilon)} \ket{\psi_1} = 2 \ket{s} + O(\epsilon) \ket{e} . \]
Solving for the initial state,
\[ \ket{s} = \sqrt{\frac{\sqrt{N} + 1}{2\sqrt{N}} + O(\epsilon)} \ket{\psi_0} + \sqrt{\frac{\sqrt{N} - 1}{2\sqrt{N}} + O(\epsilon)} \ket{\psi_1} + O(\epsilon) \ket{e} . \]

For the linear search algorithm, the system evolves by Schr\"odinger's equation \eqref{eq:Schrodinger} with a time-independent Hamiltonian $H_0$ \eqref{eq:H0}, so the state at time $t$ is
\begin{align*}
	\ket{\psi(t)}
		&= e^{-iHt} \ket{s} \\
		&= \sqrt{\frac{\sqrt{N} + 1}{2\sqrt{N}} + O(\epsilon)} e^{-iE_0t} \ket{\psi_0} + \sqrt{\frac{\sqrt{N} - 1}{2\sqrt{N}} + O(\epsilon)} e^{-iE_1t} \ket{\psi_1} + O(\epsilon) e^{-iHt} \ket{e},
\end{align*}
where the energy eigenvalues are given by \eqref{eq:2D-asymptotic}. The amplitude in basis state $\ket{m_k}$ of the $M$-dimensional subspace is
\begin{align*}
	c_k(t)
		&= \braket{m_k}{\psi(t)} \\
		&= \sqrt{\frac{\sqrt{N} + 1}{2\sqrt{N}} + O(\epsilon)} e^{-iE_0t} \braket{m_k}{\psi_0} + \sqrt{\frac{\sqrt{N} - 1}{2\sqrt{N}} + O(\epsilon)} e^{-iE_1t} \braket{m_k}{\psi_1} + O(\epsilon) \bra{m_k} e^{-iHt} \ket{e}.
\end{align*}
The inner product of $\ket{m_k}$ with $\ket{\psi_{0,1}}$ is:
\begin{align*}
	\braket{m_k}{\psi_{0,1}}
		&= \frac{1}{\sqrt{2 \frac{\sqrt{N} \pm 1}{\sqrt{N}} + O(\epsilon)}} \left( \braket{m_k}{s} \pm \braket{m_k}{w} + O(\epsilon) \braket{m_k}{e} \right) \\
		&= \frac{1}{\sqrt{2 \frac{\sqrt{N} \pm 1}{\sqrt{N}} + O(\epsilon)}} \left( \sqrt{\frac{|m_k|}{N}} \pm \delta_{k0} + O(\epsilon) \braket{m_k}{e} \right),
\end{align*}
where $\delta_{k0}$ is the Kronecker delta that equals 1 when $k = 0$ and 0 otherwise. Plugging this inner product into the amplitude,
\begin{align*}
	c_k(t)
		&= \sqrt{\frac{\frac{\sqrt{N} + 1}{2\sqrt{N}} + O(\epsilon)}{2 \frac{\sqrt{N} + 1}{\sqrt{N}} + O(\epsilon)}} e^{-iE_0t} \left( \sqrt{\frac{|m_k|}{N}} + \delta_{k0} + O(\epsilon) \braket{m_k}{e} \right) \\
		&\quad + \sqrt{\frac{\frac{\sqrt{N} - 1}{2\sqrt{N}} + O(\epsilon)}{2 \frac{\sqrt{N} - 1}{\sqrt{N}} + O(\epsilon)}} e^{-iE_1t} \left( \sqrt{\frac{|m_k|}{N}} - \delta_{k0} + O(\epsilon) \braket{m_k}{e} \right) + O(\epsilon) \bra{m_k} e^{-iHt} \ket{e}.
\end{align*}
Note that
\[ \sqrt{\frac{\frac{\sqrt{N} \pm 1}{2\sqrt{N}} + x}{2 \frac{\sqrt{N} \pm 1}{\sqrt{N}} + x}} \approx \frac{1}{2} + \frac{3x}{8} \]
for large $N$ and small $x$. Then,
\begin{align*}
	c_k(t)
		&= \left( \frac{1}{2} + O(\epsilon) \right) e^{-iE_0t} \left( \sqrt{\frac{|m_k|}{N}} + \delta_{k0} + O(\epsilon) \braket{m_k}{e} \right) \\
		&\quad + \left( \frac{1}{2} + O(\epsilon) \right) e^{-iE_1t} \left( \sqrt{\frac{|m_k|}{N}} - \delta_{k0} + O(\epsilon) \braket{m_k}{e} \right) + O(\epsilon) \bra{m_k} e^{-iHt} \ket{e}.
\end{align*}
Grouping together the exponentials
\begin{align*}
	c_k(t)
		&= \left( \frac{1}{2} + O(\epsilon) \right) \left( \sqrt{\frac{|m_k|}{N}} + O(\epsilon) \braket{m_k}{e} \right) \left( e^{-iE_0t} + e^{-iE_1t} \right) \\
		&\quad + \left( \frac{1}{2} + O(\epsilon) \right) \delta_{k0} \left( e^{-iE_0t} - e^{-iE_1t} \right) + O(\epsilon) \bra{m_k} e^{-iHt} \ket{e}.
\end{align*}
Factoring out a phase,
\begin{align*}
	c_k(t)
		= e^{-i(E_0+E_1)t/2} \Bigg[ &\left( \frac{1}{2} + O(\epsilon) \right) \left( \sqrt{\frac{|m_k|}{N}} + O(\epsilon) \braket{m_k}{e} \right) \left( e^{i\Delta E t/2} + e^{-i\Delta E t/2} \right) \\
		& + \left( \frac{1}{2} + O(\epsilon) \right) \delta_{k0} \left( e^{i\Delta E t/2} - e^{-i\Delta E t/2} \right) + O(\epsilon) \bra{m_k} e^{-iHt} \ket{e} \Bigg].
\end{align*}
Note the phase is the same for all $c_k(t)$'s, so it is a global phase and can be dropped. Also turning the sum and difference of exponentials into sines and cosines,
\[ c_k(t) = \left( 1 + O(\epsilon) \right) \left( \sqrt{\frac{|m_k|}{N}} + O(\epsilon) \braket{m_k}{e} \right) \cos \left( \frac{\Delta E t}{2} \right) + \left( 1 + O(\epsilon) \right) \delta_{k0} i\sin \left( \frac{\Delta E t}{2} \right) + O(\epsilon) \bra{m_k} e^{-iHt} \ket{e}. \]
Let us simplify the coefficient of the cosine:
\begin{align*}
	\left( 1 + O(\epsilon) \right) \left( \sqrt{\frac{|m_k|}{N}} + O(\epsilon) \braket{m_k}{e} \right)
		&= \sqrt{\frac{|m_k|}{N}} + O(\epsilon) \braket{m_k}{e} + O(\epsilon) \sqrt{\frac{|m_k|}{N}} + O(\epsilon^2) \braket{m_k}{e} \\
		&= \sqrt{\frac{|m_k|}{N}} \left( 1 + O(\epsilon) \right) + O(\epsilon) \braket{m_k}{e}.
\end{align*}
Plugging this into the amplitude,
\[ c_k(t) = \left( \sqrt{\frac{|m_k|}{N}} \left( 1 + O(\epsilon) \right) + O(\epsilon) \braket{m_k}{e} \right) \cos \left( \frac{\Delta E t}{2} \right) + \left( 1 + O(\epsilon) \right) \delta_{k0} i\sin \left( \frac{\Delta E t}{2} \right) + O(\epsilon) \bra{m_k} e^{-iHt} \ket{e}. \]
Note that
\[ O(\epsilon) \braket{m_k}{e} \cos\left[\frac{t}{\sqrt{N}}\left(1 + O(\epsilon)\right)\right] = O(\epsilon) + iO(\epsilon), \]
since $\braket{m_k}{e}$ has norm $\le 1$ and $\cos(\cdot) \le 1$. Similarly,
\[ O(\epsilon) \bra{m_k} e^{-iHt} \ket{e} = O(\epsilon) + iO(\epsilon). \]
Plugging these into the amplitude,
\[ c_k(t) = \sqrt{\frac{|m_k|}{N}} \left( 1 + O(\epsilon) \right) \cos \left( \frac{\Delta E t}{2} \right) + \left( 1 + O(\epsilon) \right) \delta_{k0} i\sin \left( \frac{\Delta E t}{2} \right) + O(\epsilon) + iO(\epsilon). \]
Factoring,
\[ c_k(t) = \left( 1 + O(\epsilon) \right) \left\{ \sqrt{\frac{|m_k|}{N}} \cos \left( \frac{\Delta E t}{2} \right) + \delta_{k0} i\sin \left( \frac{\Delta E t}{2} \right) \right\} + O(\epsilon) + iO(\epsilon). \]
Let us show that we can eliminate the factor of $(1 + O(\epsilon))$ at the beginning of the expression. First, let us focus on the real part of $c_k(t)$:
\begin{align*}
	{\rm Re}(c_k(t))
	&= \left( 1 + O(\epsilon) \right) \sqrt{\frac{|m_k|}{N}} \cos \left( \frac{\Delta E t}{2} \right) + O(\epsilon) \\
	&= \sqrt{\frac{|m_k|}{N}} \cos \left( \frac{\Delta E t}{2} \right) + \sqrt{\frac{|m_k|}{N}} \cos \left( \frac{\Delta E t}{2} \right) O(\epsilon) + O(\epsilon) \\
	&= \sqrt{\frac{|m_k|}{N}} \cos \left( \frac{\Delta E t}{2} \right) + O(\epsilon)
\end{align*}
since $\sqrt{|m_k|/N} \le 1$ and $\cos(\cdot) \le 1$. Similarly, the imaginary part simplifies:
\begin{align*}
	{\rm Im}(c_k(t))
	&= \left( 1 + O(\epsilon) \right) \delta_{k0} \sin \left( \frac{\Delta E t}{2} \right) + O(\epsilon) \\
	&= \delta_{k0} \sin \left( \frac{\Delta E t}{2} \right) + \delta_{k0} \sin \left( \frac{\Delta E t}{2} \right) O(\epsilon) + O(\epsilon) \\
	&= \delta_{k0} \sin \left( \frac{\Delta E t}{2} \right) + O(\epsilon)
\end{align*}
since $\sin(\cdot) \le 1$. So, the amplitude is
\[ c_k(t) = \sqrt{\frac{|m_k|}{N}} \cos\left[\frac{t}{\sqrt{N}}\left(1 + O(\epsilon)\right)\right] + \delta_{k0} i\sin\left[\frac{t}{\sqrt{N}}\left(1 + O(\epsilon)\right)\right] + O(\epsilon) + iO(\epsilon). \]

Let's take the norm square of this to find the probability in each basis state. The real part contributes probability
\[ {\rm Re}(c_k(t))^2 = \frac{|m_k|}{N} \cos^2 \left( \frac{\Delta E t}{2} \right) + O(\epsilon). \]
The imaginary part contributes probability
\[ {\rm Im}(c_k(t))^2 = \delta_{k0} \sin^2 \left( \frac{\Delta E t}{2} \right) + O(\epsilon). \]
Summing these, the probability in $\ket{m_k}$ is
\[ |c_k(t)|^2 = \frac{|m_k|}{N} \cos^2 \left( \frac{\Delta E t}{2} \right) + \delta_{k0} \sin^2 \left( \frac{\Delta E t}{2} \right) + O(\epsilon). \]
Plugging in $\Delta E$ from \eqref{eq:2D-asymptotic},
\[ |c_k(t)|^2 = \frac{|m_k|}{N} \cos^2\left[\frac{t}{\sqrt{N}}\left(1 + O(\epsilon)\right)\right] + \delta_{k0} \sin^2\left[\frac{t}{\sqrt{N}}\left(1 + O(\epsilon)\right)\right] + O(\epsilon). \]
Since $0\le t/\sqrt{N}\le \pi/2$ on the interval up to the first
linear peak, the functions $\cos^2 x$ and $\sin^2 x$ are Lipschitz
on the relevant bounded interval.  Thus
\[
    \cos^2\!\left[\frac{t}{\sqrt{N}}\left(1 + O(\epsilon)\right)\right]
    =
    \cos^2\!\left(\frac{t}{\sqrt{N}}\right) + O(\epsilon),
\]
and
\[
    \sin^2\!\left[\frac{t}{\sqrt{N}}\left(1 + O(\epsilon)\right)\right]
    =
    \sin^2\!\left(\frac{t}{\sqrt{N}}\right) + O(\epsilon).
\]
Using these cosine and sine approximations, $|c_k(t)|^2$ becomes
\[ |c_k(t)|^2 = \frac{|m_k|}{N} \cos^2 \left( \frac{t}{\sqrt{N}} \right) + \delta_{k0} \sin^2 \left( \frac{t}{\sqrt{N}} \right) + O(\epsilon), \]
where we used that $\epsilon$ upper bounds $|m_k|\epsilon/N$, since $|m_k| \le N$ because the number of vertices of each type can be no greater than the total number of vertices. When $k = 0$ and $k \ne 0$, we get the expressions reported in the main text in \eqref{eq:c0} and \eqref{eq:cne0}.


\section{\label{sec:cubic-proof}Proof of Sufficient Conditions for Cubic Nonlinearity on General Graphs}

We now prove the sufficient conditions for the cubic nonlinearity stated in \ref{subsubsec:sufficient-cubic}.  For later use, we first record a simple Lipschitz estimate.  Let
$P_j = \ketbra{m_j}{m_j}$ be the projector onto the $j^{\mathrm{th}}$ symmetry class.
For any two normalized states $\ket{\psi}$ and $\ket{\phi}$,
\[
\begin{aligned}
\left|
\bra{\psi}P_j\ket{\psi}
-
\bra{\phi}P_j\ket{\phi}
\right|
&=
\left|
\bra{\psi-\phi}P_j\ket{\psi}
+
\bra{\phi}P_j\ket{\psi-\phi}
\right|  \\
&\le 2\|\ket{\psi}-\ket{\phi}\|.
\end{aligned}
\]
Therefore, for the nonlinear state $\ket{\psi(t)}$ and the rescaled
linear state $\ket{\phi(t)}$,
\begin{equation}
	\label{eq:prob-nonlinear-linear-diff}
	|p_j(t) - p_j^L(s(t))|
	\le
	\frac{2\eta(t)}{|m_j|}
\end{equation}
for every class $m_j$.  Hence, for $k\ne0,\ell$,
\[
    |p_k(t) - p_\ell(t)|
    \le
    |p_k(t) - p_k^L(s(t))|
    +
    |p_k^L(s(t)) - p_\ell^L(s(t))|
    +
    |p_\ell^L(s(t)) - p_\ell(t)|.
\]
Using \eqref{eq:fne0} and \eqref{eq:fl} at $s = s(t)$, the middle term satisfies
\[
  |p_k^L(s(t)) - p_\ell^L(s(t))|
  \le
  C\epsilon
  \left(
    \frac1{|m_k|} + \frac1{|m_\ell|}
  \right),
\]
because the leading term, $\frac1N\cos^2(s(t)/\sqrt N)$, cancels.  Using this along with \eqref{eq:prob-nonlinear-linear-diff} for the first and last terms,
\[
  |p_k(t) - p_\ell(t)|
  \le
  C(\epsilon+\eta(t))
  \left(
    \frac1{|m_k|} + \frac1{|m_\ell|}
  \right).
\]
Taking the maximum over $k\ne0,\ell$, we get
\[
  \max_{k\ne0,\ell}|p_k(t) - p_\ell(t)|
  \le
  C\epsilon B_\ell + C B_\ell\eta(t),
\]
as long as $0\le s(t)\le s_*$.
Hence, \eqref{eq:Duhamel-bound-t}, \eqref{eq:cubic-remainder}, and the preceding bound give
\[
	\eta(t)
	\le
	C g\epsilon B_\ell t
	+
	C g B_\ell\int_0^t\eta(u)\,\mathrm{d}u .
\]
By Gr\"onwall's inequality~\cite{Teschl2012},
\[
	\eta(t)
	\le
	C g\epsilon B_\ell t
	\exp(CgB_\ell t).
\]
For $t = O(\sqrt{N})$, the second condition in
\eqref{eq:conditions-cubic}, namely $gB_\ell\sqrt{N}=o(1)$, keeps the
exponential factor bounded.  Thus, uniformly for $t = O(\sqrt N)$, while
$0\le s(t)\le s_*$,
\begin{equation}
	\label{eq:cubic-bootstrap-eta}
	\eta(t)
	=
	O\!\left(g\epsilon B_\ell\sqrt{N}\right)
	=
	o(1).
\end{equation}
Here the last equality follows from the second condition in \eqref{eq:conditions-cubic} and \(\epsilon=o(1)\). 
This is used in the standard continuation sense:  the estimate holds on
any interval with $s(t)\le s_*$ and $t = O(\sqrt N)$, and the lower bound
on $h(t)$ obtained next shows that the physical time to reach $s_*$ is
indeed $O(\sqrt N)$.
So, we have proved the second condition \eqref{eq:condition-eta} for the existence of the nonlinear algorithm.

Next, we estimate the time rescaling factor $h(t)$.  
For the cubic nonlinearity, $f(p) = p$, so
\[
  h(t) = 1 + g\bigl[p_0(t) - p_\ell(t)\bigr].
\]
Then,
\[
    p_0(t) - p_\ell(t)
    =
    p_0^L(s(t)) - p_\ell^L(s(t))
    +
    \bigl[p_0(t)-p_0^L(s(t))\bigr]
    -
    \bigl[p_\ell(t)-p_\ell^L(s(t))\bigr].
\]
By \eqref{eq:f0} and \eqref{eq:fl},
\[
  p_0^L(s) - p_\ell^L(s)
  =
  \sin^2\!\left(\frac{s}{\sqrt N}\right)
  +
  O(\epsilon)
  +
  O\!\left(\frac{\epsilon}{|m_\ell|}\right).
\]
Applying this with \eqref{eq:prob-nonlinear-linear-diff}, and using $|m_\ell|\ge1$, gives
\[
  p_0(t) - p_\ell(t)
  =
  \sin^2\!\left(\frac{s(t)}{\sqrt N}\right)
  +
  O(\epsilon)
  +
  O(\eta(t)).
\]
Therefore
\[
  h(t)
  =
  1 + g\sin^2\!\left(\frac{s(t)}{\sqrt N}\right)
  +
  O(g\epsilon)
  +
  O(g\eta(t)).
\]
The first condition in \eqref{eq:conditions-cubic} gives
$g\epsilon=o(1)$.  Also, by \eqref{eq:cubic-bootstrap-eta},
\[
	g\eta(t)
	=
	O\!\left(g^2\epsilon B_\ell\sqrt N\right),
\]
which is $o(1)$ by the third condition in \eqref{eq:conditions-cubic}.  Thus
\begin{equation}
	\label{eq:cubic-h-asymptotic}
	h(t)
	=
	1+g\sin^2\!\left(\frac{s(t)}{\sqrt{N}}\right)+o(1)
\end{equation}
uniformly up to the first peak.  In particular, $h(t)>0$ for all
sufficiently large $N$, satisfying the first condition \eqref{eq:condition-h} for the nonlinear algorithm to follow the rescaled linear algorithm, so $s(t)$ is monotone increasing and reaches
$s_*$ at a physical time $T_N$ satisfying \eqref{eq:runtime}.

Write the uniform error term in \eqref{eq:cubic-h-asymptotic} as
$\rho_N(t)$, with $\sup|\rho_N(t)|=\delta_N=o(1)$.  Then
\[
	h(t)
	=
	1 + g\sin^2\!\left(\frac{s(t)}{\sqrt N}\right)+\rho_N(t).
\]
For all sufficiently large $N$, $\delta_N<1/2$, and the runtime can be
sandwiched between the corresponding integrals with denominators
$1 + g\sin^2(s/\sqrt N)\pm\delta_N$.  Therefore the uniform $o(1)$ error
changes the runtime only by a relative $1+o(1)$ factor.  Using
$s_* = (\pi\sqrt N/2)[1+O(\epsilon)]$ from \eqref{eq:2D-asymptotic}, we get
\[
	T_N
	=
	\frac{\pi\sqrt{N}}{2\sqrt{1+g}}\,[1+o(1)].
\]
Finally, \eqref{eq:cubic-bootstrap-eta} and the linear success 
estimate imply that the nonlinear success probability at $T_N$ is 
$1-o(1)$.  This proves the cubic nonlinear search algorithm under the 
sufficient conditions \eqref{eq:conditions-cubic}.


\section{\label{sec:cubic-paley-proof}Proof of Sufficient Conditions for Cubic Nonlinearity on Paley Graphs}

Let $P_N$ be a Paley graph, and set
\[
	k=\frac{N-1}{2},
	\quad
	\lambda=\frac{N-5}{4},
	\quad
	\mu=\frac{N-1}{4}.
\]
In the basis $\{ \ket{m_0}=\ket{w}, \ket{m_1}, \ket{m_2} \}$,
the adjacency matrix is
\[
	A=
	\begin{pmatrix}
	0 & \sqrt{k} & 0\\
	\sqrt{k} & \lambda & \mu\\
	0 & \mu & \mu
	\end{pmatrix}.
\]
Since the Paley graph is $k$-regular, the Laplacian term differs from 
$-A/k$ only by a scalar multiple of the identity.  Thus the 
probability-equivalent linear critical Hamiltonian is
\[
	H_L=-\frac{1}{k}A-\ketbra{m_0}{m_0}.
\]
Define
\[
	\ket{r}=\frac{\ket{m_1}+\ket{m_2}}{\sqrt{2}},
	\quad
	\ket{e}=\frac{\ket{m_1}-\ket{m_2}}{\sqrt{2}},
    \quad
	n=N-1,
	\quad
	\epsilon=n^{-1/2}.
\]
In the ordered basis $(\ket{m_0},\ket{r},\ket{e})$,
\begin{equation}
	\label{eq:paley-HL-matrix}
	H_L=
	\begin{pmatrix}
	-1 & -\epsilon & -\epsilon\\
	-\epsilon & -1+\epsilon^2 & \epsilon^2\\
	-\epsilon & \epsilon^2 & \epsilon^2
	\end{pmatrix}.
\end{equation}
Standard finite-dimensional analytic perturbation 
theory~\cite{Kato1995} gives, uniformly for 
$0\le \sigma\le \sigma_*:=\pi/(2\epsilon)$,
\begin{equation}\label{eq:paley-linear-asymptotics}
\begin{aligned}
	a_L(\sigma)&:=\bra{m_0} e^{-iH_L\sigma} \ket{s}
	=e^{i\sigma}\left[\epsilon\cos(\epsilon\sigma)+i\sin(\epsilon\sigma)\right]+O(\epsilon^2), \\
	r_L(\sigma)&:=\langle r | e^{-iH_L\sigma} | s \rangle
	=e^{i\sigma}\left[\cos(\epsilon\sigma)+i\epsilon\sin(\epsilon\sigma)\right]+O(\epsilon^2), \\
	e_L(\sigma)&:=\langle e | e^{-iH_L\sigma} | s \rangle
	=i\epsilon e^{i\sigma}\sin(\epsilon\sigma)+O(\epsilon^2).
\end{aligned}
\end{equation}
Indeed, the two relevant eigenvalues are 
$-1\mp\epsilon+O(\epsilon^3)$, so the $O(\epsilon^2)$ errors 
in \eqref{eq:paley-linear-asymptotics} remain uniform for times 
$\epsilon\sigma = O(1)$.

Writing the amplitude in $\ket{m_1}$ and $\ket{m_2}$ as
\[
	b_L(\sigma)=\frac{r_L(\sigma)+e_L(\sigma)}{\sqrt{2}},
	\quad
	c_L(\sigma)=\frac{r_L(\sigma)-e_L(\sigma)}{\sqrt{2}},
\]
the total probability imbalance between the two unmarked classes in 
the linear evolution is
\[
	Q_L(\sigma) = |b_L(\sigma)|^2-|c_L(\sigma)|^2
	= 2\operatorname{Re}\!\left[r_L(\sigma)\overline{e_L(\sigma)}\right].
\]
The leading term in $r_L\overline{e_L}$ is purely imaginary, so
\begin{equation}
	\label{eq:paley-Qlin}
	Q_L(\sigma)=O(\epsilon^2)=O(N^{-1})
\end{equation}
uniformly for $0\le\sigma\le\sigma_*$.  Also,
\begin{equation}
	\label{eq:paley-p0lin}
	|a_L(\sigma)|^2
	=
	\sin^2(\epsilon\sigma)+\epsilon^2\cos^2(\epsilon\sigma)+O(\epsilon^2).
\end{equation}

Now let $\ket{\psi(t)} = a(t)\ket{m_0}+b(t)\ket{m_1}+c(t)\ket{m_2}$ be the 
nonlinear Paley evolution with cubic nonlinearity, $g = N-1 = 2k$, and 
$\ell=1$.  From \eqref{eq:R} and \eqref{eq:gammac},
\begin{equation}
	\label{eq:paley-decomp-cubic}
	i\frac{d}{dt}\ket{\psi(t)}
	=
	\left[h(t)H_L+R(t)\right]\ket{\psi(t)},
\end{equation}
where, exactly,
\begin{align}
	Q(t) &= |b(t)|^2-|c(t)|^2, \nonumber \\
	R(t) &= 2Q(t)\ketbra{m_2}{m_2}, \label{eq:paley-h-R-Q} \\
	h(t) &= N|a(t)|^2-Q(t). \nonumber
\end{align}
Let
\[
	\sigma(t)=\int_0^t h(u)\,\mathrm{d}u,
	\quad
	\ket{\phi(t)} = e^{-iH_L\sigma(t)}\ket{s},
\]
and define
\[
	\eta(t) = \|\ket{\psi(t)} - \ket{\phi(t)}\|.
\]
For any fixed $T$, Duhamel's formula \eqref{eq:Duhamel-bound-t} and $R(t)$ from \eqref{eq:paley-h-R-Q} imply, 
on the interval where $0\le t\le T$ and $0\le\sigma(t)\le\sigma_*$,
\[
	\eta(t)
	\le
	\int_0^t \|R(u)\|\,\mathrm{d}u
	=
	2\int_0^t |Q(u)|\,\mathrm{d}u .
\]
The map $\ket{\psi}\mapsto Q = |b|^2-|c|^2$ is Lipschitz on the unit 
sphere, so
\[
	|Q(u) - Q_L(\sigma(u))|\le 4\eta(u),
\]
or
\[
    |Q(u)| \le |Q_L(\sigma(u))| + 4\eta(u).
\]
Plugging this into the above inequality from Duhamel's formula,
\begin{align*}
    \eta(t) 
        &\le 2\int_0^t |Q(u)|\,\mathrm{d}u \\
        &\le 2\int_0^t \bigl(|Q_L(\sigma(u))| + 4\eta(u)\bigr) \,\mathrm{d}u \\
        &\le C \int_0^t \frac{1}{N} \,\mathrm{d}u + 8 \int_0^t \eta(u) \,\mathrm{d}u \\
        &= \frac{Ct}{N} + 8\int_0^t\eta(u)\,\mathrm{d}u,
\end{align*}
where in the third line, we used \eqref{eq:paley-Qlin}.
By Gr\"onwall's inequality \cite{Teschl2012},
\begin{equation}
	\label{eq:paley-eta-bound}
	\sup_{0\le t\le T,\,0\le\sigma(t)\le\sigma_*}
	\eta(t)
	=
	O_T(N^{-1}),
\end{equation}
where the subscript $T$ means that the hidden constant may depend on the
fixed value of $T$, but not on $N$ or on $t\in[0,T]$.
Consequently, on every bounded physical-time interval before the first 
peak,
\begin{equation}
	\label{eq:paley-Q-bound}
	Q(t) = O_T(N^{-1})
\end{equation}
and
\begin{equation}
    \label{eq:paley-a-squared}
	|a(t)|^2 = |a_L(\sigma(t))|^2+O_T(N^{-1}).
\end{equation}

For the bounded-time argument, fix $T_0=2\pi$ and use the estimates above
with $T=T_0$.  The argument below shows that the hitting time satisfies
$T_N<T_0$, so this use of the $O_T$ bounds is self-consistent.  We next show that the rescaled time reaches the linear peak in bounded 
physical time.  Taking $h(t)$ from \eqref{eq:paley-h-R-Q}, we plug in
\eqref{eq:paley-Q-bound} and \eqref{eq:paley-a-squared} to get
\[
  h(t) = N|a_L(\sigma(t))|^2+O_{T_0}(1)
\]
whenever $0\le \sigma(t)\le\sigma_*$.  Using
\eqref{eq:paley-p0lin}
and writing $n = N-1 = \epsilon^{-2}$, we obtain
\[
\begin{aligned}
  N|a_L(\sigma)|^2
  &=
  (n+1)\sin^2(\epsilon\sigma)
  +
  \left(1+\frac1n\right)\cos^2(\epsilon\sigma)
  +
  O(1)\\
  &=
  1 + n\sin^2(\epsilon\sigma)+O(1).
\end{aligned}
\]
Therefore
\begin{equation}
    \label{eq:paley-h-global}
    h(t) = 1 + n\sin^2(\epsilon\sigma(t)) + O_{T_0}(1).
\end{equation}

This global estimate is not sharp enough near $\sigma=0$, where the main
term is only order one.  On any fixed interval $0\le\sigma\le\sigma_0$,
we instead Taylor expand the linear amplitude $a_L(\sigma)$ in \eqref{eq:paley-linear-asymptotics} using $\cos(\epsilon\sigma) \approx 1$ and $\sin(\epsilon\sigma) \approx \epsilon\sigma$, resulting in
\[
  a_L(\sigma)
  =
  e^{i\sigma}\epsilon(1+i\sigma)+O_{\sigma_0}(\epsilon^2).
\]
Since $a(t) = a_L(\sigma(t))+\delta a(t)$ and 
$|\delta a(t)|\le
\|\ket{\psi(t)}-\ket{\phi(t)}\| = O_{T_0}(\epsilon^2)$,
we have
\[
\begin{aligned}
  |a(t)|^2 - |a_L(\sigma(t))|^2
  &=
  2\operatorname{Re}\!\left(
  \overline{a_L(\sigma(t))}\,\delta a(t)
  \right)
  +
  |\delta a(t)|^2 .
\end{aligned}
\]
On the interval $0\le\sigma(t)\le\sigma_0$, we have
$a_L(\sigma(t)) = O_{\sigma_0}(\epsilon)$, so
\[
  |a(t)|^2 - |a_L(\sigma(t))|^2
  =
  O_{T_0,\sigma_0}(\epsilon^3).
\]
Multiplying by $N = O(\epsilon^{-2})$, the nonlinear correction to
$N|a|^2$ is $O_{T_0,\sigma_0}(\epsilon)$.  On the same interval,
\[
	N|a_L(\sigma)|^2
	=
	1 + \sigma^2 + O_{\sigma_0}(\epsilon),
\]
and \eqref{eq:paley-Q-bound} gives
\[
	Q(t) = O_{T_0}(N^{-1}) = O_{T_0}(\epsilon^2).
\]
Using the exact identity $h(t) = N|a(t)|^2 - Q(t)$ from
\eqref{eq:paley-h-R-Q}, we obtain
\[
  h(t) = 1 + \sigma(t)^2 + O_{T_0,\sigma_0}(\epsilon)
\]
for $0\le\sigma(t)\le\sigma_0$.
Since $h(t)$ converges to $1 + \sigma(t)^2$, for all sufficiently large $N$, $h(t)$ will be at least half this value, i.e.,
\begin{equation}
	\label{eq:paley-h-lower-small}
	h(t)\ge \frac{1}{2}\bigl[1+\sigma(t)^2\bigr]
	\qquad (0\le\sigma(t)\le\sigma_0).
\end{equation}
Let $C_{T_0}$ bound the $O_{T_0}(1)$ term in
\eqref{eq:paley-h-global}.  Using
$\sin x\ge 2x/\pi$ for $0\le x\le \pi/2$, choose $\sigma_0$
large enough that
\[
    1 + \frac{4\sigma_0^2}{\pi^2}\ge 2C_{T_0}.
\]
Then, for $\sigma_0\le\sigma\le\sigma_*$, the main term
$1 + n\sin^2(\epsilon\sigma)$ dominates the $O_{T_0}(1)$ error.  Then
\begin{equation}
	\label{eq:paley-h-lower-large}
	h(t)
	\ge
	\frac{1}{2}\left[1+n\sin^2(\epsilon\sigma(t))\right]
	\qquad (\sigma_0\le\sigma(t)\le\sigma_*).
\end{equation}
Since $\mathrm{d}\sigma/\mathrm{d}t = h(t)>0$, the physical time needed to reach 
$\sigma_*$ is at most
\[
	2\int_0^{\sigma_0}\frac{\mathrm{d}\sigma}{1+\sigma^2}
	+
	2\int_{\sigma_0}^{\sigma_*}
	\frac{\mathrm{d}\sigma}{1+n\sin^2(\epsilon\sigma)}.
\]
The first integral is $2\arctan(\sigma_0) < \pi$, and the second is 
bounded by
\[
	2\sqrt{n}\int_0^{\pi/2}\frac{\mathrm{d}x}{1+n\sin^2 x}
	=
	\pi\sqrt{\frac{n}{n+1}}
	<\pi,
\]
where we used the standard integral
\[
	\int_0^{\pi/2}\frac{\mathrm{d}x}{1+n\sin^2 x}
	=
	\frac{\pi}{2\sqrt{1+n}}.
\]
Therefore, there is a time $T_N < 2\pi = T_0$ such that $\sigma(T_N) = \sigma_*$.    
At $\sigma_* = \pi/(2\epsilon)$, we have
\[
  \cos(\epsilon\sigma_*) = 0,
  \quad
  \sin(\epsilon\sigma_*) = 1.
\]
Thus \eqref{eq:paley-linear-asymptotics} gives
\[
  a_L(\sigma_*)
  =
  i e^{i\sigma_*} + O(\epsilon^2).
\]
Taking the squared modulus,
\[
  |a_L(\sigma_*)|^2
  =
  1 + O(\epsilon^2)
  =
  1 + O(N^{-1}).
\]
Since probabilities are at most $1$, this implies
\[
  |a_L(\sigma_*)|^2
  =
  1 - O(N^{-1})
\]
in the usual success probability sense.  Since $T_N < T_0$, we may apply
\eqref{eq:paley-eta-bound} with $T = T_0$, giving
\[
	|a(T_N)-a_L(\sigma_*)| = O(N^{-1}).
\]
Therefore
\[
	|a(T_N)|^2 = 1-O(N^{-1}).
\]
Thus the cubic nonlinear search algorithm on Paley graphs with $g = N-1$ 
succeeds with probability $1 - O(N^{-1})$ in bounded physical time.  


\section{\label{sec:cubic-quintic-proof}Proof of Sufficient Conditions for Cubic-Quintic Nonlinearity on General Graphs}

As shown in \eqref{eq:cq-remainder-bound}, the cubic-quintic remainder is bounded by the same quantity that appeared in the cubic case in
\eqref{eq:cubic-remainder}.  Hence, the bootstrap estimate in Appendix~\ref{sec:cubic-proof} leading to
\eqref{eq:cubic-bootstrap-eta} applies without change to the
cubic-quintic remainder.  In particular, under the conditions
\eqref{eq:conditions-cubic}, the nonlinear cubic-quintic trajectory stays
$o(1)$-close to the corresponding rescaled linear trajectory up to the
first success time, provided $0\le s(t)\le s_*$ and $t = O(\sqrt N)$.

The time rescaling factor is different from the cubic case.  For the
cubic-quintic nonlinearity,
\[
    h(t) = 1 + g\left[f(p_0(t))-f(p_\ell(t))\right].
\]
Using \eqref{eq:cq-difference-identity},
\begin{equation}
    \label{eq:cq-h-factorization}
    f(p_0(t))-f(p_\ell(t))
    =
    \bigl[p_0(t)-p_\ell(t)\bigr]
    \bigl[1-p_0(t)-p_\ell(t)\bigr].
\end{equation}
We estimate the two factors using the linear estimates and the
nonlinear-linear comparison from the cubic subsection.  From
\eqref{eq:f0}, \eqref{eq:fl}, and
\eqref{eq:prob-nonlinear-linear-diff},
\begin{equation}
    \label{eq:cq-diff-p0-pl}
    p_0(t)-p_\ell(t)
    =
    \sin^2\!\left(\frac{s(t)}{\sqrt N}\right)
    +O(\epsilon)+O(\eta(t)).
\end{equation}
Similarly,
\begin{equation}
    \label{eq:cq-sum-p0-pl}
    p_0(t)+p_\ell(t)
    =
    \sin^2\!\left(\frac{s(t)}{\sqrt N}\right)
    +O\!\left(\frac1N\right)
    +O(\epsilon)+O(\eta(t)).
\end{equation}
Indeed, the extra $O(1/N)$ term in \eqref{eq:cq-sum-p0-pl} comes from
adding the two copies of
$(1/N)\cos^2(s/\sqrt N)$ in \eqref{eq:f0} and \eqref{eq:fl}.  Combining
\eqref{eq:cq-h-factorization}--\eqref{eq:cq-sum-p0-pl}, we obtain
\[
    f(p_0(t))-f(p_\ell(t))
    =
    \sin^2\!\left(\frac{s(t)}{\sqrt N}\right)
    \cos^2\!\left(\frac{s(t)}{\sqrt N}\right)
    +O\!\left(\frac1N\right)
    +O(\epsilon)
    +O(\eta(t)),
    \label{eq:cq-f-difference-asymptotic}
\]
where the error is uniform up to the first peak.  Hence
\[
    h(t)
    =
    1
    +g\sin^2\!\left(\frac{s(t)}{\sqrt N}\right)
      \cos^2\!\left(\frac{s(t)}{\sqrt N}\right)
    +O\!\left(\frac{g}{N}\right)
    +O(g\epsilon)
    +O(g\eta(t)). \label{eq:cq-h-asymptotic-pre}
\]
Under \eqref{eq:conditions-cubic}, we have $g\epsilon = o(1)$ and,
using \eqref{eq:cubic-bootstrap-eta},
\[
    g\eta(t) = O\!\left(g^2\epsilon B_\ell\sqrt N\right) = o(1).
\]
For the graph families to which the analytical conclusion below is
applied, $g/N = o(1)$ as well.  Therefore
\begin{equation}
    \label{eq:cq-h-asymptotic}
    h(t)
    =
    1
    +g\sin^2\!\left(\frac{s(t)}{\sqrt N}\right)
      \cos^2\!\left(\frac{s(t)}{\sqrt N}\right)
    +o(1)
\end{equation}
uniformly up to the first peak.  In particular, $h(t)>0$ for all
sufficiently large $N$, satisfying \eqref{eq:condition-h}, so $s(t)$ is monotone increasing and reaches
$s_*$ at a physical time $T_N$ satisfying \eqref{eq:runtime}.

Substituting \eqref{eq:cq-h-asymptotic} into \eqref{eq:runtime} gives
\[
    T_N
    =
    \int_0^{s_*}
    \frac{\mathrm{d}s}{
    1
    +g\sin^2(s/\sqrt N)\cos^2(s/\sqrt N)
    +o(1)
    }.
\]
Using $s_* = (\pi\sqrt N/2)[1+O(\epsilon)]$ from
\eqref{eq:2D-asymptotic} and the change of variables $x = s/\sqrt N$, the
uniform $o(1)$ term in the denominator changes the integral by only a
relative $1+o(1)$ factor, and we get
\[
    T_N
    =
    \sqrt N
    \int_0^{\pi/2}
    \frac{\mathrm{d}x}{1+g\sin^2x\cos^2x}
    \,[1+o(1)].
\]
Since
\[
    \int_0^{\pi/2}
    \frac{\mathrm{d}x}{1+g\sin^2x\cos^2x}
    =
    \frac{\pi}{2\sqrt{1+g/4}},
\]
we obtain
\[
    T_N
    =
    \frac{\pi\sqrt N}{2\sqrt{1+g/4}}\,[1+o(1)],
\]
as reported in \eqref{eq:cq-runtime-modest}.
Finally, the accumulated-remainder estimate and the linear success
estimate imply that the nonlinear success probability at this time is
$1-o(1)$.

\end{widetext}


\bibliography{refs}

\end{document}